# FMR-related phenomena in spintronic devices

Yi Wang, Rajagopalan Ramaswamy, and Hyunsoo Yang

*Department of Electrical and Computer Engineering, National University of Singapore, 117576, Singapore*

Spintronic devices, such as non-volatile magnetic random access memories and logic devices, have attracted considerable attention as potential candidates for future high efficient data storage and computing technology. In a heavy metal or other emerging material with strong spin-orbit coupling (SOC), the charge currents induce spin currents or spin accumulations via SOC. The generated spin currents can exert spin-orbit torques (SOTs) on an adjacent ferromagnet, which opens up a new way to realize magnetization dynamics and switching of the ferromagnetic layer for spintronic devices. In the SOT scheme, the charge-to-spin interconversion efficiency (SOT efficiency) is an important figure of merit for applications. For the effective characterization of this efficiency, the ferromagnetic resonance (FMR) based methods, such as the spin transfer torque ferromagnetic resonance (ST-FMR) and the spin pumping, are common utilized in addition to low frequency harmonic or dc measurements. In this review, we focus on the ST-FMR measurements for the evaluation of the SOT efficiency. We provide a brief summary of the different ST-FMR setups and data analysis methods. We then discuss ST-FMR and SOT studies in various materials, including heavy metals and alloys, topological insulators, two dimensional (2D) materials, interfaces with strong Rashba effect, antiferromagnetic materials, two dimensional electron gas (2DEG) in oxide materials and oxidized nonmagnetic materials.

# CONTENTS

# 1. Introduction

In the recent years, the electrical manipulation of the magnetization is a central theme for modern spintronics [1-3]. Following the successful applications of magnetic tunnel junctions (MTJs) [4, 5] in the read heads of hard disk drives, new functional spintronic devices, such as non-volatile magnetic random access memories (MRAM) and magnetic logic devices [6, 7], are being pursued enthusiastically by researchers nowadays. For practical spintronic applications, it is desired to achieve efficient magnetization manipulation via electrical means with a higher operation speed and lower energy consumption. In order to evaluate the efficiency of the electrical manipulation of the magnetization as well as to understand underlying physics in the magnetic devices, various techniques based on magnetization dynamics under application of currents have been developed.

The study of magnetization dynamics can be dated back to the observation of ferromagnetic resonance (FMR) in the first half of the $20^{th}$ century [8, 9]. Originally, FMR was used to study magnetic films and the magnetization dynamics in such films in the context of external magnetic fields only. With the discovery of spin transfer torques (STT) in the year 1996 [10, 11] and development of the modern nano-fabrication techniques which allow the device dimensions to be scaled-down, electrical manipulation of nanomagnets using spin-polarized current injection gained popularity. After the year 2000, the STT driven magnetization switching using electric currents was experimentally observed in giant magnetoresistance (GMR) nanopillars and nanosized MTJ devices [12-14]. In order to get insight into the magnetization dynamics of an individual nanomagnet or magnetic device under the application of an electric current, the spin transfer torque ferromagnetic resonance (ST-FMR) technique was developed and explored in GMR nanopillars and nanosized MTJs [15-18].

One of the drawbacks of the STT scheme is the requirement of an additional magnet to create spin polarized currents. In the year 2004, the imaging of direct charge-to-spin current conversion in a single semiconductor GaAs due to the spin Hall effect (SHE) was observed [19]. Ever since this observations, pure spin current generation in non-magnets has drawn tremendous attention for spintronic devices [2]. Recently, it was demonstrated that the magnetization can be switched or driven into precession modes by only pure spin currents [20-24], which opened up a new route to realize spintronic devices. In order to achieve highly efficient magnetization manipulation, a key challenge is to identify materials that exhibit a large charge-to-spin interconversion efficiency (i.e. spin Hall angle or spin-orbit torque efficiency). The ST-FMR is an effective technique to evaluate the spin-orbit torque (SOT) efficiency. The first demonstration was reported in a Pt/NiFe bilayer planar spin Hall device in 2011 [25], which is based on the spin rectification of the anisotropic magnetoresistance (AMR) effect. Subsequently, ST-FMR has been used to evaluate the SOT efficiencies and understand the underlying physics in many emerging materials as presented later on.

This article provides a topical review of ST-FMR technique and its applications in studying the SOT related phenomena in diverse emerging materials. In Section 2, we begin with the basic principle of magnetization dynamics. In Section 3 and 4, we introduce the ST-FMR setups and data analysis methods. In Section 5, 6 and 7, we review the progress in the SOT studies by using ST-FMR technique in different material systems, such as heavy metals, topological insulators and other emerging material systems, and we end with a summary and perspectives in Section 8.

## 2. Magnetization dynamics

### 2.1. LLG equation

Under macrospin approximation, the time domain dynamics of the magnetization vector (**m**)

of a ferromagnet in the presence of an effective magnetic field ($\mathbf{H}_{eff}$) can be described by the Landau-Lifshitz-Gilbert (LLG) equation [26],

$$\frac{\partial \mathbf{m}}{\partial t} = \left(-\gamma\mu_0 \mathbf{m} \times \mathbf{H}_{eff}\right) + \left(\alpha \mathbf{m} \times \frac{\partial \mathbf{m}}{\partial t}\right) \quad (1)$$

where $\gamma$, $\mu_0$, and $\alpha$ are the gyromagnetic ratio, vacuum permeability, the Gilbert damping parameter, respectively. The right hand side of the LLG equation has two torque terms – the first term, $-\gamma\mu_0 \mathbf{m} \times \mathbf{H}_{eff}$, is a precession term that rotates $\mathbf{m}$ around $\mathbf{H}_{eff}$, while the second term, $\alpha \mathbf{m} \times \frac{\partial \mathbf{m}}{\partial t}$, is the damping (or energy dissipation) term which aligns $\mathbf{m}$ along $\mathbf{H}_{eff}$. The value of $\alpha$ determines the rate at which $\mathbf{m}$ damps towards the $\mathbf{H}_{eff}$. When the value of $\alpha$ is large (small), $\mathbf{m}$ damps towards $\mathbf{H}_{eff}$ at a faster (slower) rate.

When a spin polarized current (polarized along $\hat{\boldsymbol{\sigma}}$) interacts with $\mathbf{m}$, there are additional torques experienced by $\mathbf{m}$ and the LLG equation is modified as [10, 11, 27]

$$\frac{\partial \mathbf{m}}{\partial t} = -\gamma\mu_0 \mathbf{m} \times \mathbf{H}_{eff} + \alpha \mathbf{m} \times \frac{\partial \mathbf{m}}{\partial t} + \boldsymbol{\tau}_{DL} + \boldsymbol{\tau}_{FL} \quad (2)$$

where $\boldsymbol{\tau}_{DL}$ is of the $\mathbf{m} \times (\hat{\boldsymbol{\sigma}} \times \mathbf{m})$ symmetry and is called the damping-like torque, while $\boldsymbol{\tau}_{FL}$ is of the $\hat{\boldsymbol{\sigma}} \times \mathbf{m}$ symmetry and is called the field-like torque.

## 2.2. Ferromagnetic resonance (FMR)

In Section 2.1, it was discussed that in the presence of $\mathbf{H}_{eff}$, $\mathbf{m}$ precesses about $\mathbf{H}_{eff}$. Due to energy dissipation (damping) in the ferromagnet, the amplitude of these precessions continuously decays. An external energy supplied to the ferromagnet in the form of an alternating magnetic field ($H_{RF}$) can compensate this energy dissipation leading to forced precession of $\mathbf{m}$. If the frequency of $H_{RF}$ is tuned to the natural precession frequency of $\mathbf{m}$, the amplitude of the forced precession will be maximum due to resonance. This resonance is called the ferromagnetic resonance and the precession frequency during the resonance condition is called the resonance frequency ($f=\omega_{res}/2\pi$). $f$ and $H_0$ (the resonant field) are related by the Kittel's relation,

$$f = (\gamma/2\pi)[H_0(H_0 + 4\pi M_{eff})]^{1/2} \tag{3}$$

where $4\pi M_{eff}$ is the effective demagnetization field of the ferromagnet. According to equation (3), the resonance condition can be matched by fixing $H_0$ and tuning the frequency of $H_{RF}$ or vice versa.

## 2.3. Spin transfer torque ferromagnetic resonance (ST-FMR)

In contrast to the magnetic field driven FMR, it is possible to excite FMR using spin torques generated by an alternating current, called ST-FMR. The ST-FMR technique was first used in a Pt/NiFe (Py) bilayer in the case of a spin Hall scheme [25], which is proved to be an effective method to evaluate the charge-to-spin conversion efficiency (i.e. spin Hall angle). We first provide a more general description of ST-FMR process in the Pt/Py bilayer as an example.

As shown in figure 1(a), when an in-plane rf charge current ($I_{RF}$) is injected into the Pt/Py bilayer, non-equilibrium spins are generated at the Pt/Py interface due to the spin-orbit coupling (SOC) in the Pt layer. Subsequently, these generated spins diffuse into the adjacent Py layer and exert SOTs (damping-like torque, $\tau_{DL}$ and/or field-like torque, $\tau_{FL}$) on the Py local magnetization. In addition, the current in the Pt layer induced Oersted field can also exert the Oersted field torque ($\tau_{Oe}$) on the Py magnetization. Since the injected charge current is oscillatory, all the above torques are oscillatory and hence trigger the Py magnetization into precession around the direction of effective magnetic field, which leads to an oscillatory AMR in the ST-FMR device. Consequently, the oscillation of the AMR and the $I_{RF}$ in the ST-FMR device produce a mixing ST-FMR voltage $V_{mix}$ across the Pt/Py bilayer, which can be detected by a dc voltmeter or a lock-in amplifier. Since the ST-FMR signal is the spin rectification of the magnetoresistance in the device [28], the ST-FMR technique has a high enough sensitivity and can be used to study the microsized or even nanosized magnetic devices.

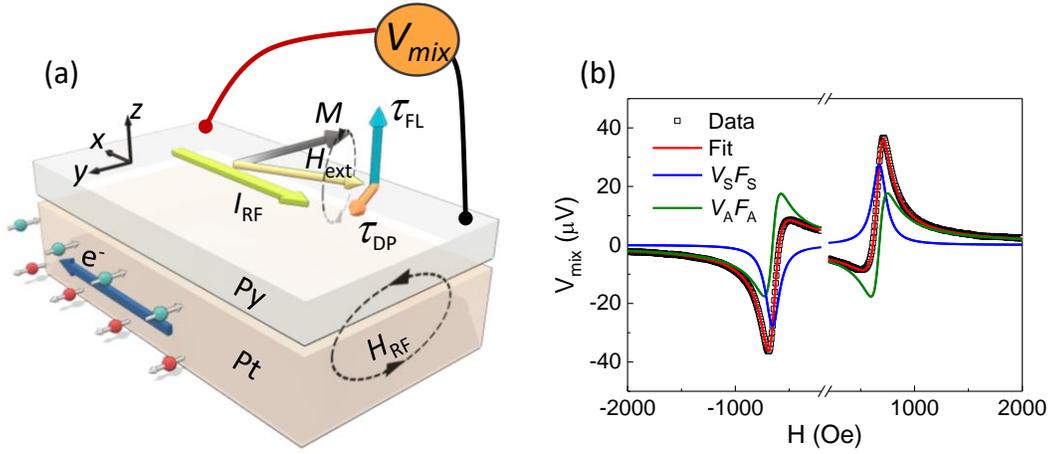

**Figure 1.** (a) Illustration of spin current generation in Pt and SOT driven Py magnetization dynamics in the ST-FMR measurements. The big blue arrow denotes the electron flow direction (opposite to the $I_{RF}$ direction). The blue with green and red balls denote the generated spin polarizations in the Pt layer. (b) A typical ST-FMR signal (open symbols) from a Pt (6 nm)/Py (4 nm) device at 6 GHz with fits (solid curves), where the blue and green curves represent the symmetric Lorentzian ($V_S F_S$) and antisymmetric Lorentzian ($V_A F_A$) components, respectively.

Subsequently, we present a derivation of general expression for the ST-FMR signal $V_{\mathrm{mix}}$. The applied rf current to the Pt/Py ST-FMR device is written as $I_{RF} = I\cos(\omega t)$, where $I$ is the amplitude of $I_{RF}$ with a frequency of $\omega/2\pi$. The device resistance is $R(t) = R_0 + \Delta R \cos^2\theta(t)$, where $R_0$ is the resistance when $I_{RF}$ is perpendicular to the Py magnetization direction, $\Delta R$ is the AMR. $\theta(t)$ is the angle between $I_{RF}$ and Py magnetization, which is a function of time $t$ due to magnetization precession, given by the equation $\theta(t) = \theta_H + \theta_c \cos(\omega t + \delta)$, where $\theta_H$ is the constant angle between $H_{ext}$ and $I_{RF}$, $\theta_c$ is the cone angle of Py magnetization precession, which is a measure of the total torques on the Py magnetization. Usually, $\theta_c$ is much smaller than $\theta_H$. $\delta$ is the resonance phase between the driven force (i.e. damping-like torque $\tau_{DL}$, field-like torque $\tau_{FL}$ or Oersted field torque $\tau_{Oe}$) and the magnetization response (i.e. the oscillation of AMR) [29, 30].

By using the Taylor's expansion, we get $\cos\theta(t) = \cos\theta_H - \sin\theta_H \cdot \theta_c \cos(\omega t + \delta)$ and $R(t) = R_0 + \Delta R[\cos^2\theta_H - 2\cos\theta_H \sin\theta_H \cdot \theta_c \cos(\omega t + \delta)]$. Hence, the product is

$$V(t) = I(t)R(t)$$
$$= (IR_0 + I\Delta R\cos^2\theta_H)\cos(\omega t) - I\Delta R\sin(2\theta_H)\cdot\theta_c\cos(2\omega t + \delta)/2$$
$$- I\Delta R\sin(2\theta_H)\cdot\theta_c(\cos\delta)/2$$

The total voltage $V(t)$ comprises of three terms at frequencies $\omega$, $2\omega$, and a time-independent term. The last dc term is the rectified ST-FMR signal, $V_{mix} = -I\Delta R\sin(2\theta_H)\cdot\theta_c(\cos\delta)/2$. The ST-FMR signal is determined by the combination of the amplitude of $I_{RF}$, AMR in the device, the angle of $H_{ext}$ with respect to $I_{RF}$, the cone angle of magnetization precession $\theta_c$ and the resonance phase $\delta$.

For an in-plane alternative magnetic field driven magnetization dynamics of an in-plane magnetization [30, 31], the value of $\delta$ changes from $\pi$ to 0 around the resonance field $H_0$ with a linewidth of $\Delta H$ (full width at half maximum) and $\delta = \pi/2$ at the resonance field $H_0$. Therefore, the $V_{mix}$ attributed to the in-plane alternative magnetic field has an antisymmetric Lorentzian line shape, which is the case for the $\tau_{Oe}$ and/or $\tau_{FL}$ driven magnetization dynamics in a Pt/Py ST-FMR device as shown by the green curves in figure 1(b). Whereas for the $\tau_{DL}$ driven magnetization dynamics, there is an additional 90° phase difference compared to $\tau_{FL}$ or $\tau_{Oe}$ driven magnetization dynamics for an AMR based rectification system. Therefore, the driving force $\tau_{DL}$ is in-phase ($\delta = 0$) at the resonance field $H_0$, leading to a symmetric Lorentzian line shape arising from the damping-like torque in a Pt/Py ST-FMR device as illustrated by the blue curves in figure 1(b). In Section 4, we will present three quantitative analyses of the ST-FMR signal in detail.

## 3. ST-FMR setups

### 3.1. Traditional ST-FMR setup with a dc voltmeter

As described in Section 2.3, the ST-FMR signal is a rectified dc voltage, which can be detected directly by a dc voltmeter. Figure 2 shows a schematic of the traditional ST-FMR setup with a dc voltmeter. It depicts a typical ST-FMR device with a micro-strip of heavy metal (HM)/ferromagnet (FM) bilayer structure. A ground-signal-ground (G-S-G) coplanar waveguide (CPW) is fabricated with one G and one S electrodes contacting the HM/FM bilayer channel for applying an $I_{RF}$ to the ST-FMR device which is supplied by a signal generator (SG) connected to an rf port of a bias tee. A ST-FMR signal can be simultaneously detected by a voltmeter, such as Keithley 2182A and Keithley 2002, connected to the low frequency port of the bias tee as shown in the figure 2. The traditional ST-FMR setup with a dc voltmeter usually works well for the devices having considerable amplitude of ST-FMR signals with the value of several tens of µV, such as Pt/Py systems [25, 32].

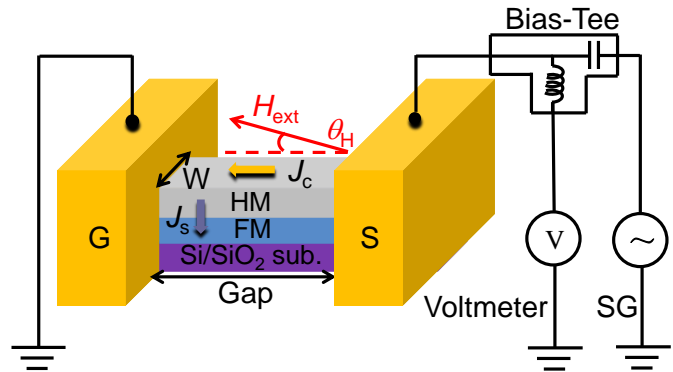

**Figure 2.** The schematic of the traditional ST-FMR measurement setup, illustrating the ST-FMR device and measurement circuit including a bias tee, a high frequency signal generator and a dc voltmeter. Adapted from [32], with the permission of AIP Publishing.

### 3.2. Precise ST-FMR setup with a lock-in amplifier

For the traditional ST-FMR setup, the noise level is usually in the order of ~100 nV because of the limitation of a dc voltmeter. This noise level reduces the signal-to-noise ratio substantially while studying some of the emerging material systems possessing smaller ST-FMR signals with the order of ~µV, such as the topological insulator/FM bilayers [33, 34]. In order to decrease the noise (i.e., increase the signal-to-noise ratio), a more sensitive ST-FMR setup was developed in which the dc voltmeter is replaced with a lock-in amplifier [16, 35]. Figure 3(a) shows the ST-FMR setup with lock-in technique.

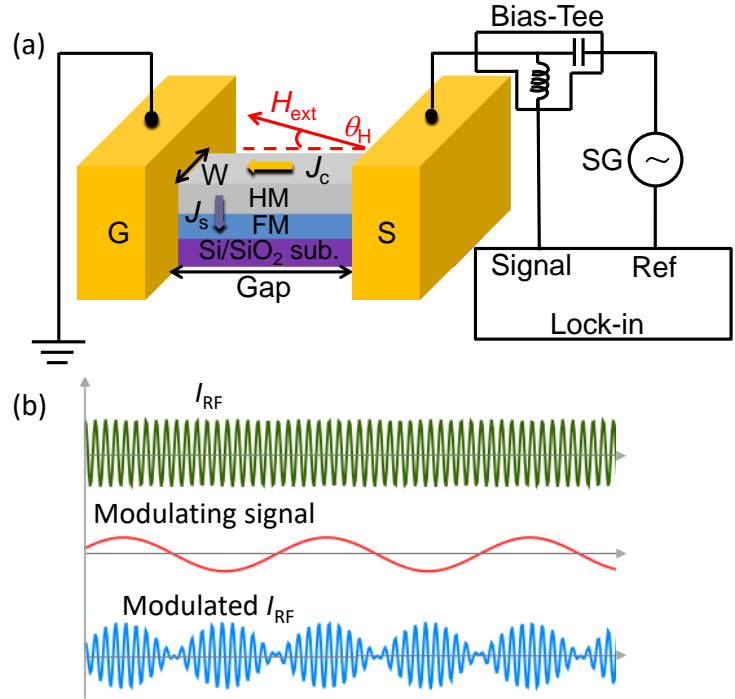

**Figure 3.** (a) The schematic of the sensitive ST-FMR setup with a lock-in amplifier, illustrating the ST-FMR device and measurement circuit including a bias tee, a high frequency signal generator and a lock-in amplifier. Adapted from [32], with the permission of AIP Publishing. (b) Illustration of the amplitude modulation of input $I_{RF}$ from a signal generator by a modulating signal with a much lower frequency, and the modulated $I_{RF}$ is finally injected into the ST-FMR device.

The input $I_{RF}$ is amplitude modulated by a low frequency sinusoidal wave signal (modulating signal) as illustrated in figure 3(b). The low frequency modulating signal is also provided as a reference signal into the reference port of the lock-in amplifier and thus serves as a trigger. The modulated $I_{RF}$ is injected into the ST-FMR device to give rise to the ST-FMR signal. The ST-FMR signal, in this case, is a low frequency voltage (instead of a dc) since the $I_{RF}$ is amplitude modulated. The low frequency ST-FMR signal can be detected by the lock-in amplifier with the phase locking technique. Therefore, the noise level can be less than ~10 nV, thus providing a better signal-to-noise ratio compared to the traditional ST-FMR setup.

### 3.3. Design of CPW and impedance match

In order to avoid a large amount of rf power loss or reflection during the rf current transmission in the CPW and the HM/FM current channel, the ST-FMR device should be designed to have an impedance ($Z_m$) of around 50 Ω. First, the CPW should be designed considering impedance matching. One example of a typical CPW is illustrated in figure 4(a). The width of S is $s$ and the gap between S and G line is $w$. The characteristic impedance of the CPW can be tuned by varying the $s$ and/or $w$ values. One can use a common CPW calculator software to select the proper combination of $s$ and $w$ values for the CPW design. For example, the $s$ is about 60 μm and $w$ is about 30 μm for the CPW design in Ref. [32]. Further, the HM/FM bilayer current channel should also be impedance matched, which can be tuned by changing the channel length $L$ and width $W$ (also see the current channel denoted by the red circles in the figures 4(b) and (c)). For example, $L$ is in the range of ~10 – 60 μm and $W$ is in the range of ~10 – 30 μm in Refs. [32-34]. Finally, $Z_m$ of ST-FMR devices can be quantitatively calibrated by the microwave-network-analysis measurement using a vector network analyzer (VNA).

Usually, there are two kinds of CPW designs, namely the asymmetric and the symmetric CPW designs. The asymmetric CPW design is shown in figure 4(b), which was initially used for the HM/FM bilayer ST-FMR measurements [25]. However, there is a limitation of the asymmetric CPW. To elaborate on this limitation, consider the example of Pt/Py ST-FMR device as shown in figure 4(d). In addition to the current induced in-plane Oersted field ($H_{RF}$) to the Py magnetization, an imbalanced current induced out-of-plane Oersted field $H_z$ from the contacting electrodes can be produced as $I_{RF}$ is transmitting along the CPW [32, 36]. Consequently, $H_z$ exerts an in-plane torque ($-m \times H_z$) on the Py magnetization and generates symmetric voltage signals for both directions of $H_{ext}$, which superimpose onto the symmetric signals arising from damping-like torque, $\tau_{DL}$, from spin currents. Therefore, the $H_z$ induced torque might lead to an overestimation or underestimation of the spin Hall angle in Pt. However, one can eliminate this contamination by carefully locating the GSG probe position in the middle of the CPW [32].

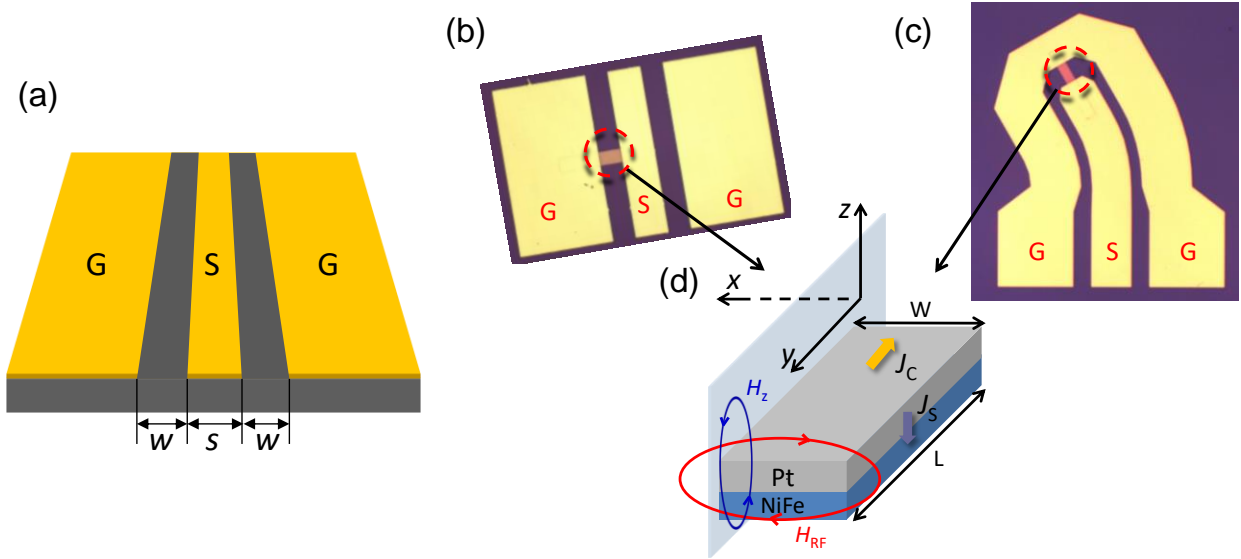

**Figure 4.** (a) The schematic of a typical CPW, illustrating the ground-signal-ground (G-S-G) transmission line. ST-FMR devices with (b) asymmetric CPW and (c) symmetric CPW. (d) The Pt/Py microstrip with the charge current $J_C$, spin current $J_S$, current induced in-plane Oersted field ($H_{RF}$). $H_z$ is the current induced imbalanced out-of-plane Oersted field in the ST-FMR device with asymmetric CPW design. Adapted from [32], with the permission of AIP Publishing.

In order to diminish or eliminate the imbalanced $H_z$ and the related influences in the ST-FMR measurements and spin Hall angle evaluation, the symmetric CPW design was developed. As shown in figure 4(c), the current induced Oersted field in the electrodes is balanced as $I_{RF}$ flows in the symmetric CPW, resulting in no $H_z$ component. Therefore, the symmetric CPW design is better than the asymmetric CPW for ST-FMR measurements.

**4. Methods to evaluate the spin-orbit torque (SOT) and SOT efficiency**

In Section 2.3, a general derivation of ST-FMR signal was provided. In this section, we will further present three quantitative analyses of the ST-FMR signal. The ST-FMR signals from a HM/FM bilayer device (see figure 1(b)) can be fitted by $V_{mix} = V_S F_S(H_{ext}) + V_A F_A(H_{ext})$, where $F_S(H_{ext}) = \Delta H^2/[(H_{ext} - H_0)^2 + \Delta H^2]$ is a symmetric Lorentzian function of amplitude $V_S$ and $F_A(H_{ext}) = \Delta H(H_{ext} - H_0)/[(H_{ext} - H_0)^2 + \Delta H^2]$ is an antisymmetric Lorentzian function of amplitude $V_A$ [25, 32]. The symmetric component has a maximum value at the resonance field $H_0$ and is centered about $H_0$, whereas, the antisymmetric component has a dispersive curve with the value of zero at $H_0$. Both components have the same linewidth $\Delta H$ with respect to the magnetic field.

### 4.1. Ratio of $V_S/V_A$

It is noted that $V_S$ is proportional to the amplitude of spin currents $J_S$ (i.e. $\tau_{DL}$) and is written as $V_S \propto \hbar J_S/(2e\mu_0 M_s t)$ [18, 25, 37], where $M_s$ and $t$ are the saturation magnetization and thickness of the FM layer, respectively, and $\hbar$ is the reduced Planck constant. While $V_A$ is correlated with $H_{RF}$ induced Oersted field torque ($\tau_{Oe}$), which is written as $V_A \propto H_{RF}[1 + (4\pi M_{eff}/H_{ext})]^{1/2}$ [18, 25, 37], where $M_{eff}$ is the effective magnetization of the FM layer. Note that the field-like torque $\tau_{FL}$ is assumed to be negligible here. $H_{RF}$ is estimated by Ampere's law as $H_{RF} = J_C d/2$, where $J_C$ is the charge current density in the HM layer. Note that the charge current flowing in the FM layer is assumed to be spatially uniform, therefore, there is no net Oersted field torque on the FM magnetization itself.

In the spin Hall scheme, the charge-to-spin conversion efficiency (i.e. spin Hall angle), defined as $\theta_{sh} = J_S/J_C$, can be found from the ratio of $V_S/V_A$ [25]. Using the above equations, one obtains the spin Hall angle as $\theta_{sh} = (V_S/V_A)(e\mu_0 M_s t d/\hbar)[1 + (4\pi M_{eff}/H_{ext})]^{1/2}$. This method is self-calibrated in the sense that the torques on the FM magnetization only arise from $J_C$ in the Pt layer and the spin Hall angle can be easily calculated without knowing the exact values of $I_{RF}$ and $H_{RF}$ in ST-FMR devices. The method works well to determine $\theta_{sh}$ under the assumption that the $V_A$ only attributes to $H_{RF}$. In other words, this method may give rise to a wrong estimate of $\theta_{sh}$ for the case of a non-negligible field-like torque $\tau_{FL}$, which might arise due to the interfacial effects, such as the Rashba effect. Since $\tau_{FL}$ can also produce an antisymmetric Lorentzian line shape signal similar to $H_{RF}$, the value of $\theta_{sh}$ might be overestimated or underestimated from the method of $V_S/V_A$ [38, 39]. Therefore, one should be aware of no significant field-like torque from interfacial effect before using the analysis method of the ratio of $V_S/V_A$.

Note that for simplicity, hereafter, we refer the spin Hall angle as the SOT efficiency, unless otherwise specified, in order to represent the charge-to-spin conversion efficiency in a variety of materials. After obtaining the SOT efficiency, we can easily evaluate the damping-like torque using $\tau_{DL} = \theta_{sh} J_C \hbar/(2eM_s t)$.

**4.2. Only-$V_S$**

Since some material systems give rise to a significant $\tau_{FL}$ to the adjacent FM magnetization, the SOT efficiency can be better determined by analyzing the symmetric component $V_S$ only [33, 34, 40]. The damping like torque, $\tau_{DL}$, and the associated in-plane SOT efficiency, $\theta_{sh}$, can be evaluated using $V_S = -\frac{I_{RF} \gamma \cos\theta_H}{4} \frac{dR}{d\theta_H} \tau_{DL} \frac{1}{\Delta} F_S(H_{ext})$, $\sigma_S = J_S/E = \tau_{DL} M_s t/E$, and $\theta_{sh} = \sigma_S/\sigma$ [33, 34, 40], where $dR/d\theta_H$ is the angular dependent magnetoresistance at $\theta_H$, $\Delta$ is the linewidth of ST-FMR signal in the frequency domain, $E$ is the microwave electric field across the ST-FMR device, and $\sigma_S$ and $\sigma$ are the spin Hall and longitudinal charge conductivities of the HM layer, respectively. In addition, the field-like torque, $\tau_{FL}$, can be derived by $V_A = -\frac{I_{rf} \gamma \cos\theta_H}{4} \frac{dR}{d\theta_H} (\tau_{FL} + \tau_{Oe}) \frac{[1+(\mu_0 M_{eff}/H_{ext})]^{1/2}}{\Delta} F_{asym}(H_{ext})$ [33, 40]. In order to obtain $\tau_{FL}$, the $\tau_{Oe}$ should be evaluated separately and excluded [33]. Similarly, the out-of-plane SOT efficiency, $\theta_\perp$, can be evaluated accordingly.

Since the resistance of the HM/FM microstrip is written as $R = R_0 + \Delta R \cos^2(\theta_H)$ due to the AMR effect (see Section 2.3), we obtain $\frac{dR}{d\theta_H} = -2\Delta R \sin(\theta_H)\cos(\theta_H) \propto \sin(\theta_H)\cos(\theta_H)$. Consequently, we find $V_{mix} \propto \cos^2(\theta_H)\sin(\theta_H)$ from the relation, $V_{mix} \propto \frac{dR}{d\theta_H}\cos(\theta_H)$ [32]. Thus, it is observed that even though the change of AMR is maximum at $\theta_H = 45°$, the largest ST-FMR signals are achieved at $\theta_H \sim 35°$.

The comparison of the in-plane SOT efficiencies evaluated from the ratio of $V_S/V_A$ method and from only-$V_S$ method is presented in Section 5.1 for Pt and Section 5.2 for Ta. These results suggest that the only-$V_S$ method is a more general way to evaluate the SOT efficiency in materials with SOC. However, compared to the $V_S/V_A$ ratio method, the only-$V_S$ technique requires additional measurements to determine the $dR/d\theta_H$ and $\sigma$, and also requires quantitative determination of $I_{RF}$ through a ST-FMR device by considering the device impedance and rf power losses using the microwave-network analysis measurement [32].

Note that the SOT efficiency or in-plane SOT efficiency in the entire review for a variety of materials is denoted as $\theta_{sh}$, and the out-of-plane SOT efficiency is denoted as $\theta_\perp$, in order to be consistent.

### 4.3. Modulation of damping (MOD)

In the context of conventional STT, researchers found that the STT and magnetization fluctuation can be modulated in the vertical MTJs by applying a dc current bias [41, 42]. Recently, this method was used for HM/FM bilayer systems in the spin Hall scheme and the magnetization dynamics can be tuned by applying a dc current in the HM layer [43]. Based on the STT theory [41], the effective magnetic damping ($\alpha$) and thus the FMR linewidth ($\Delta H$) will increase or decrease depending on the relative direction of spin polarizations with respect to the magnetization direction. The relationship can be written as [25]

$$\Delta H = (\Delta H)_0 + (\Delta H)_{sh} = \frac{2\pi f}{\gamma}\alpha + \frac{2\pi f}{\gamma}\frac{\sin\theta_H}{(H_{ext}+2\pi M_{eff})\mu_0 M_s t}\frac{\hbar}{2e}J_C\theta_{sh},$$ where $(\Delta H)_0$ is the linewidth

at zero dc bias, $(\Delta H)_{sh}$ is the modulated linewidth due to the dc bias induced spin currents, and $f$ is the rf current frequency. Therefore, one can perform the ST-FMR measurements to extract the $\Delta H$ as a function of $J_C$ as shown in figure 5. By linearly fitting the data, the SOT efficiency is obtained accordingly.

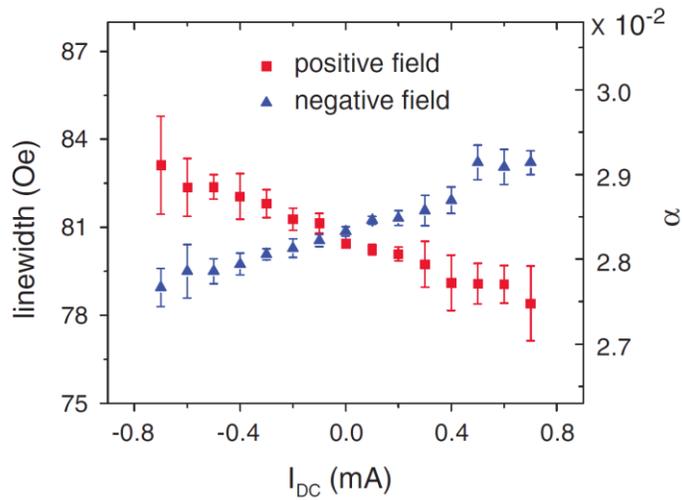

**Figure 5.** The FMR linewidth and Gilbert damping coefficient $\alpha$ as a function of dc current, $I_{dc}$, in the Pt (6 nm)/Py (4 nm) bilayer device at $f = 8$ GHz. Reprinted with permission from [25]. Copyright (2011) by the American Physical Society.

The modulation of damping (MOD) method is an extension of the ST-FMR measurements. So far, it has been used to evaluate the SOT efficiency commonly in Pt and Ta [25, 44-47], which usually have considerably large ST-FMR signals. Recently, it has been utilized for the SOT efficiency evaluation of metallic antiferromagnetic materials [48]. However, for the material systems possessing smaller ST-FMR signals, it is challenging to employ the MOD method to evaluate the SOT efficiency [40].

Table 1 summarizes the advantages and disadvantages of the above three SOT efficiency evaluation methods. The method of $V_S/V_A$ is simple, however it works only for the material systems where the $V_A$ only attributes to $H_{RF}$ (i.e. there is no significant field-like torque). Therefore, this method has limitations for the SOT efficiency evaluation. However, the method of only-$V_S$ can successfully address this issue and can be a more general method to extract the SOT efficiency in a wide range of materials.

| Methods | $V_S/V_A$ | MOD | Only-$V_S$ |
| --- | --- | --- | --- |
| Parameters required to estimate $\theta_{sh}$ | $V_S$, $V_A$, $M_s$, $t$, $d$, $M_{eff}$ and $H_{ext}$ | $(\Delta H)_{sh}$, $f$, $\theta_H$, $H_{ext}$, $M_s$, $t$, $M_{eff}$, and $J_C$ | $V_S$, $I_{RF}$, $\theta_H$, $dR/d\theta_H$, $\Delta$, $M_s$, $t$, $E$ and $\sigma$ |
| Analysis procedure | Can directly get $\theta_{sh}$ from ST-FMR lineshape analysis | Needs precise ST-FMR linewidth measurements and analysis | Needs additional AMR ($dR/d\theta_H$), conductivity ($\sigma$) measurements and rf power analysis |
| Material systems | Limited to material systems without significant $\tau_{FL}$ | Material systems with considerably large and clean ST-FMR signals | Almost all material systems showing ST-FMR signals |
| Remarks | A simple method but with serious limitations | A less simple method but with moderate limitations | A more general method, but need extra measured parameters |
| References | [25], [32] | [25], [44-48] | [32-34], [40] |

Table 1. Summary of the advantages and disadvantages of three SOT efficiency evaluation methods.

## 5. Spin Hall effect (SHE) in heavy metals and alloys

Spin Hall effect [49-52] is an electrical technique, which exploits the bulk SOC in a nonmagnetic material (such as HM) to convert charge currents into pure spin currents without any external magnetic field. The first theoretical predictions of SHE by Dyakonov and Perel dates back to 1971 [53] and, in 1999, the SHE method to generate spin currents was reinvigorated by Hirsch [49]. Subsequently, in 2004, the SHE was first experimentally observed in the GaAs system [19]. The phenomenon of SHE relies on the spin-dependent asymmetric scattering of the unpolarized electrons due to the bulk SOC in the nonmagnetic materials to generate a pure transversal spin current density. The SOC effects that give rise to SHE in a nonmagnetic material may be due to the band structure of the material (intrinsic SHE) or by addition of high SOC impurities into the material (extrinsic SHE). As illustrated in figure 6, the spin polarization of the accumulated spins due to SHE is orthogonal to both the directions of the injected charge current as well as the spin current. Thus, the spin current generation using SHE can be described using the equation $J_S = \theta_{sh}(J_C \times \hat{\sigma})$, where $\hat{\sigma}$ is the spin polarization. Here, $\theta_{sh} (= J_S/J_C)$ is spin Hall angle of the nonmagnetic material and quantifies the efficiency of the spin current generation by SHE. For further details into SHE, the readers can refer to detailed reviews on SHE [51, 52].

For practical spintronic applications, it is desirable to have a higher spin Hall angle for efficient spin current generation using SHE. It is generally understood that the in-plane SOT efficiency in the ST-FMR technique estimates the spin Hall angle of the NM. Recently, many efforts have been devoted to obtain a large SOT efficiency and efficient SOT driven magnetization switching or precession via pure spin currents in HM/FM bilayers [2, 20-24]. Apart from ST-FMR measurements [25, 32], several different techniques have been also employed to determine the spin Hall angle, such as the lateral spin valve method [54-56], spin pumping [57] and spin Hall magnetoresistance (SMR) measurements [58]. In the following sub-sections, we first review the research progress of the SOT efficiency evaluation in HMs and alloys by the ST-FMR technique.

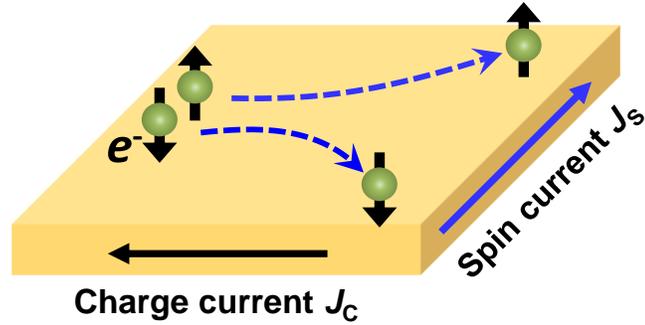

**Figure 6.** Schematic of SHE. The black-colored, blue-colored, and dotted arrows indicate the directions of charge current $J_C$, spin current $J_S$, and the motions of spin-up and spin down electrons.

### 5.1. Spin Hall effect in Pt

Liu *et al.* first utilized ST-FMR to evaluate the SOT efficiency in the Pt (6 nm)/Py (4 nm) bilayer device at room temperature [25]. The schematic of a Pt/Py bilayer and the main results are shown in figure 7. The value of SOT efficiency in Pt is determined to be ~0.056. Further, they have also utilized the MOD method as an independent check to confirm their results. By taking advantage of the large SOT efficiency, they also demonstrated the SOT driven magnetization switching in Pt/Co/Al$_2$O$_3$ trilayer structures [22].

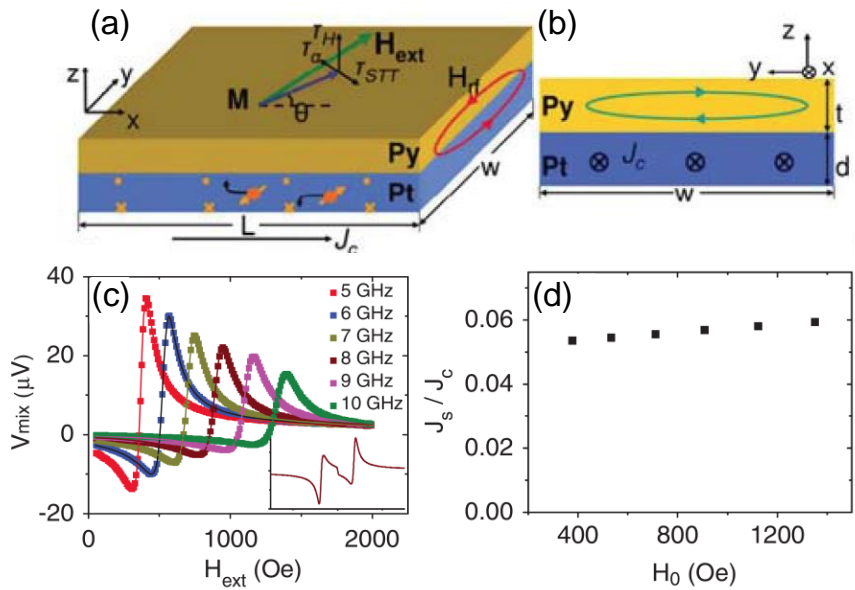

**Figure 7.** (a) Schematic of a Pt (6 nm)/Py (4 nm) bilayer illustrating the spin-orbit torque $\tau_\text{STT}$, the Oersted field torque $\tau_\text{H}$, and the direction of the damping torque $\tau_\alpha$. (b) Left-side view of the Pt/Py system, showing the Oersted field generated by the current flowing just in the Py layer, which should produce no net effect on the Py magnetization. (c) ST-FMR signals on a Pt/Py sample. The sample dimensions are 20 µm wide and 110 µm long. The inset shows the ST-FMR signal of 8 GHz in a full magnetic field range. (d) SOT efficiency in Pt/Py at different resonance fields. Reprinted with permission from [25]. Copyright (2011) by the American Physical Society.

Subsequently, the SOT efficiency in Pt has been quantified by different measurement methods in different research groups. However, there have been significant disagreements in the reported values of the SOT efficiency in Pt, ranging from ~0.012 to ~0.12 [32, 51, 52]. To figure out the possible reasons for this discrepancy and determine the intrinsic SOT efficiency value in Pt is of great importance. For ST-FMR measurements, the measured SOT efficiency is evaluated from the amount of spins that are absorbed by the FM layer. There are three aspects that affect the measured SOT efficiency. The first one is the spin diffusion in the Py layer since the spins can transmit into Py with a characteristic length (i.e. spin diffusion length), which is often ignored. Wang *et al.* [32] carried out Py thickness dependent ST-FMR measurements in Pt (6 nm)/Py ($t = 2 - 10$ nm) bilayers. It was found that $\theta_\text{sh}$ increases when the Py thickness increases as shown in figure 8(a). By taking into account the spin diffusion in the Py layer, the SOT efficiency of Pt is determined to be ~0.068 at room temperature [32]. The second aspect is spin diffusion in the Pt layer. $J_\text{S}$ and the measured SOT efficiency $\theta_\text{sh}$ in a Pt film of thickness $d$ should be reduced from the bulk value ($d = \infty$) by $J_\text{S}(d) = J_\text{S}(\infty)\left[1 - \text{sech}(d/\lambda_\text{S})\right]$ and $\theta_\text{sh} = \theta_\infty\left[1 - \text{sech}(d/\lambda_\text{S})\right]$, respectively [25, 32, 45, 59]. For example, the spin diffusion length, $\lambda_\text{S}$ of Pt estimated from a Pt/Py device is ~1.5 nm [32] (figure 8(b)). The third aspect is the interface transparency as spin currents transmit through the Pt and Py interface, which is discussed in Section 5.4.

Figure 8(c) shows a summary of the SOT efficiencies as a function of $\lambda_\text{S}$ in Pt measured by different techniques from various research groups [25, 32, 58, 60-69]. A clear correlation between SOT efficiency $\theta_{sh}$ and $\lambda_S$ is found, which is approximately an inverse relationship with $\theta_{sh}\lambda_S \sim$ 0.13 nm (denoted by the blue thick line) [32]. The $\lambda_S$ in a material is generally proportional to its electrical conductivity ($\sigma$) [32, 70]. Figure 8(d) shows an approximately linear relationship between the reported $\lambda_S$ and $\sigma$ in Pt films [32, 45, 55, 58-62, 64, 67, 71]. Therefore, the SOT efficiency is inversely related to the conductivity of the HMs. All the data in figures 8(c) and (d) were measured at room temperature, except for those from Refs. [61] and [71] which were measured at 10 K, as denoted by small black stars. In addition, it was found that the $\theta_{sh}$ in Pt remains almost constant as temperature decreases from 300 to 13 K [32, 55].

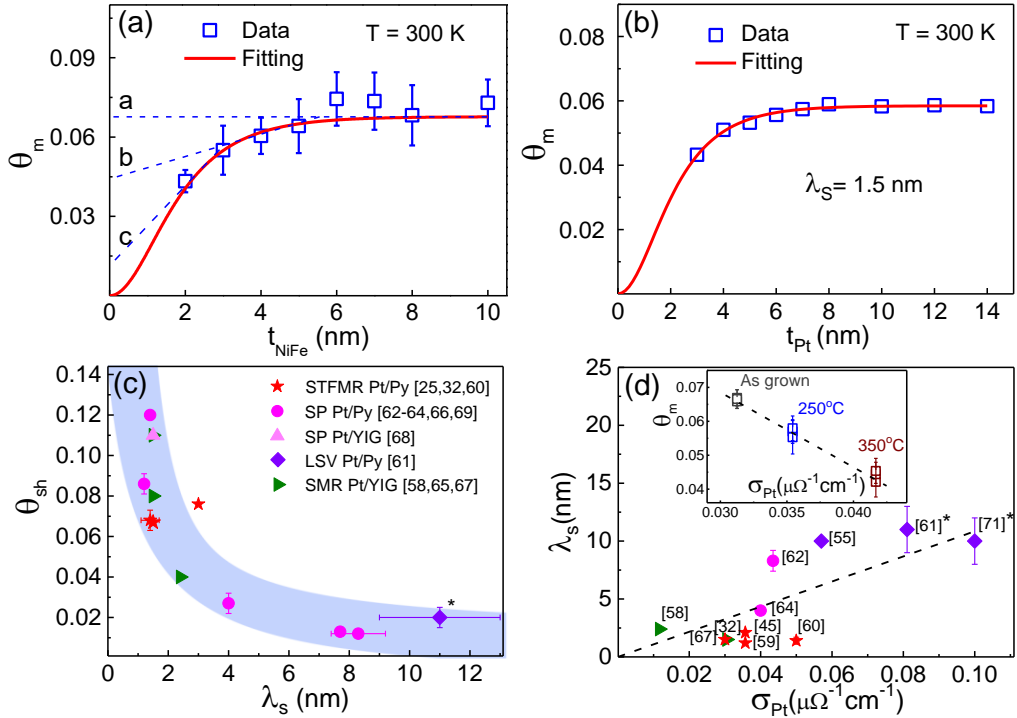

**Figure 8.** (a) Measured SOT efficiency $\theta_{sh}$ for Pt/Py (*t*) devices with different Py thicknesses from ST-FMR measurements at 8 GHz and 300 K. (b) $\theta_{sh}$ (blue squares) of Pt (*d*)/Py (4 nm) devices as a function of Pt thickness at 300 K. (c) Summarized SOT efficiency $\theta_{sh}$ as a function of $\lambda_S$ in Pt from various reports. The ST-FMR, SP, LSV, and SMR represent spin torque ferromagnetic resonance, spin pumping, nonlocal measurement in lateral spin valves, and spin Hall

magnetoresistance method, respectively. The blue thick curve shows a correlation $\theta_{sh}\lambda_S \sim 0.13$ nm. (d) $\lambda_S$ in Pt as a function of its electrical conductivity ($\sigma_{Pt}$). The inset shows the measured SOT efficiency in various Pt films with different conductivities. The dashed lines serve as guide for the eye. Adapted from [32], with the permission of AIP Publishing.

It is possible that in ST-FMR measurements, spin pumping can also occur due to the inverse spin Hall effect by $J_C = \theta_{sh}(J_S \times \hat{\sigma})$. Therefore, an additional dc voltage ($V_{SP}$) can also be produced and might contaminate the ST-FMR signals. The spin pumping contributions for Pt (6 nm)/Py ($t =$ 2 – 10 nm) bilayers have been estimated in Ref. [32], by using the following equations in the method part in Ref. [40].

$$V_{SP} = \theta_{sh}\frac{eW\lambda_S R}{2\pi}\tanh(\frac{d}{2\lambda_S})\text{Re}(g_{\uparrow\downarrow}^{\text{eff}})\omega(\theta_c)^2\sin(\theta_H)\sqrt{H_0/(H_0 + 4\pi M_{\text{eff}})},$$

$$\theta_c = \frac{1}{dR/d\theta_H}\frac{2}{I_{RF}}\sqrt{(V_S)^2 + (V_A)^2},$$

where $W$ is the channel width, $R$ is the device resistance, $d$ is the Pt thickness, Re($g_{\uparrow\downarrow}^{\text{eff}}$) is the real part of the effective spin mixing conductance ($\approx 2\times10^{19}$ m$^{-2}$), $\theta_c$ is the maximum precession cone angle in the device plane, and $V_S$ and $V_A$ are the symmetric and antisymmetric components of the ST-FMR signal, respectively. It was found that the estimated spin pumping signals $V_{sp}$ are at least one order of magnitude smaller than the ST-FMR symmetric component $V_S$. It is known that $V_{sp}$ is proportional to device resistance [57]. Since the resistance of the ST-FMR device is usually small (~50 Ω for the impedance matching, the spin pumping contributions in the ST-FMR measurements are usually much smaller than the ST-FMR signals.

In addition, the SOT efficiency in Pt/Py bilayers has been quantified by the $V_S/V_A$ ratio and only-$V_S$ methods separately [32], and no clear difference between the SOT efficiencies from these two methods was found. This suggests that there is negligible $\tau_{FL}$ arising from the Pt/Py interface.

However, it has been also reported that $\tau_{FL}$ might arise from the Rashba effect at the Pt/FM interface [20, 38, 72], which indicates that the characteristics of the interfaces, such as the interface quality and the degree of oxidation of the interfaces, could also affect the SOT efficiency. On the other hand, a Cu with negligible SOC or other insertion layers can be inserted between the HM and FM layer to eliminate or modify the interface SOT effect [46, 73, 74].

**5.2. Spin Hall effect in Ta**

In 2012, a large SOT efficiency and giant spin Hall induced magnetization switching were reported in a Ta (8 nm)/Co$_{40}$Fe$_{40}$B$_{20}$ (CoFeB, 4 nm) bilayer at room temperature [21]. Figures 9(a) and (b) illustrate the ST-FMR measurements and results in a Ta/CoFeB bilayer. A very high SOT efficiency of ~0.15 was determined by the $V_S/V_A$ ratio method. The SOT efficiency shows a negative sign compared to the positive sign in Pt, which agrees with the opposite character of spin accumulation in Pt and Ta. Moreover, an efficient SOT driven magnetization switching was first realized in a three terminal MTJ device (Ta/CoFeB/MgO/CoFeB), which opens a new avenue to the SOT based spintronic devices.

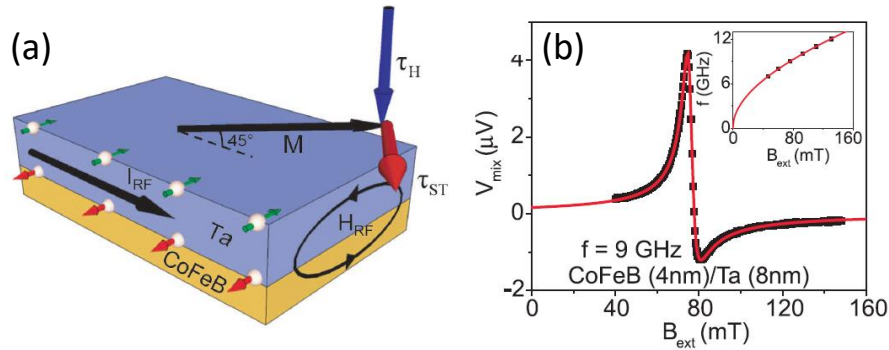

**Figure 9.** (a) Schematic of sample geometry for a Ta (8 nm)/CoFeB (4 nm) bilayer ST-FMR device. (b) ST-FMR signal at $f$ = 9 GHz. From [21]. Reprinted with permission from AAAS.

The interface effect between the Ta and CoFeB layer has not been considered by Liu *et al.* [21] in the Ta SOT efficiency evaluation. Thus, we compare the SOT efficiency in Ta on Si/SiO$_2$

substrate/Co$_{40}$Fe$_{40}$B$_{20}$ (4 nm)/Ta (8 nm) and Si/SiO$_2$ substrate/Py (4 nm)/Ta (8 nm) devices by ST-FMR measurements at different temperatures. A significant contribution of field-like torque, $\tau_{FL}$, is observed, which might arise from the interface between Ta and CoFeB layers. Figure 10(a) shows ST-FMR signals and fits (solid lines) in a CoFeB (4 nm)/Ta (8 nm) device for frequencies between 6–10 GHz at room temperature. The $H_{ext}$ is swept at an angle of $\theta_H = 35°$ with respect to the $I_{RF}$. The SOT efficiency in Ta is quantified by the $V_S/V_A$ ratio and only-$V_S$ methods, respectively. As shown in figure 10(b), the in-plane SOT efficiency ($\theta_{sh}$) in CoFeB/Ta device is ~0.109 from the $V_S/V_A$ ratio method at room temperature and it remains in the range of ~0.1–0.15 at most of temperatures. The $\theta_{sh}$ value is similar to the previous report [21]. In addition, we perform the only-$V_S$ method to determine $\theta_{sh}$ as well as the out-of-plane SOT efficiency ($\theta_\perp$) which characterizes the $\tau_{FL}$. Surprisingly, the $\theta_{sh}$ value from the only-$V_S$ method is ~0.02 at room temperature as shown by the red circles in figure 10(b). This big inconsistency in $\theta_{sh}$ indicates a significant $\tau_{FL}$ (i.e. $\theta_\perp$) contribution in the CoFeB/Ta bilayer. As shown in figure 10(c), there is indeed a large $\theta_\perp$ of ~0.044 in a CoFeB/Ta bilayer at room temperature. This result is qualitatively similar to an earlier report of a larger $\tau_{FL}$ in CoFeB/Ta compared to $\tau_{DL}$ [75].

Similarly, we also characterize the Py (4 nm)/Ta (8 nm) bilayer device. As shown in figure 10(d), the $\theta_{sh}$ from the only-$V_S$ method is ~0.01 at room temperature. The small inconsistency in $\theta_{sh}$ from two analysis methods indicates that the $\tau_{FL}$ is not significant in the Py/Ta bilayer. Finally, $\theta_{sh}$ in Ta is ~0.01–0.02 obtained from the FM/Ta bilayer using ST-FMR at room temperature, which is consistent with other reports [61, 76-78]. The slightly different $\theta_{sh}$ from Py/Ta and CoFeB/Ta can be attributed to the different interface transparency [76, 79-81]. The $\theta_{sh}$ (from only-$V_S$) almost remains constant from 15 to 300 K, while the $\theta_\perp$ shows a fast decrease at low temperature range, suggesting that $\theta_{sh}$ and $\theta_\perp$ might come from different origins. Moreover, we observed a similar $\theta_{sh}$ of ~0.01–0.015 in Py (20 nm)/Ta (6–25 nm) and CoFeB (15 nm)/Ta (15 nm) bilayers by spin pumping measurements at room temperature. This similar SOT efficiencies from both ST-FMR and spin pumping measurements suggest the Onsager reciprocity, which has been also demonstrated recently [82].

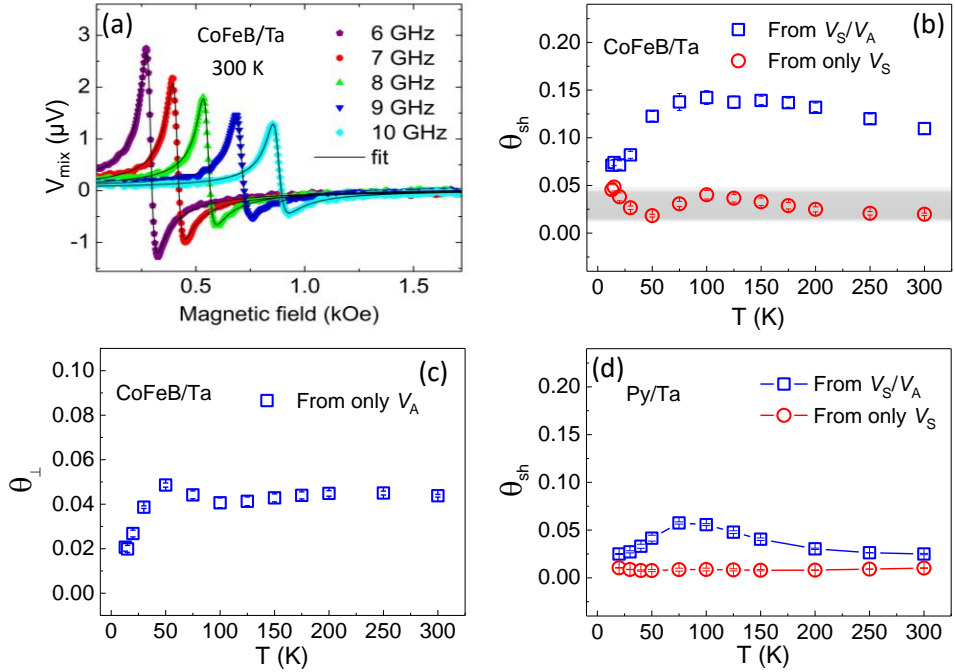

**Figure 10.** (a) ST-FMR measurements in a CoFeB (4 nm)/Ta (8 nm) device at different frequencies. The symbols are the measured data, and the solid lines are fits by sum of Lorentzian functions. (b) The in-plane SOT efficiency and (c) the out-of-plane SOT efficiency in Ta as a function of temperature obtained from ST-FMR measurements on the CoFeB (4 nm)/Ta (8 nm) device. (d) The in-plane SOT efficiency as a function of temperature in the Py (4 nm)/Ta (8 nm) device.

Figure 11 shows a summary of the SOT efficiencies in Ta films, ranging from 0.0037 to 0.26, measured by different techniques from different research groups [61, 75-77, 83-89]. The fit of $\theta_{sh}$ as a function of $\lambda_S$ indicates an approximately inverse relationship with $\theta_{sh}\lambda_S \sim 0.036$ nm (denoted by the blue thick line). However, it is observed that the correlation between $\theta_{sh}$ and $\lambda_S$ is not as clear as the case of Pt as shown in figure 8(c), which might be due to multi-phases (such as α-, β-phases or amorphous film) in different Ta thin films [21, 87].

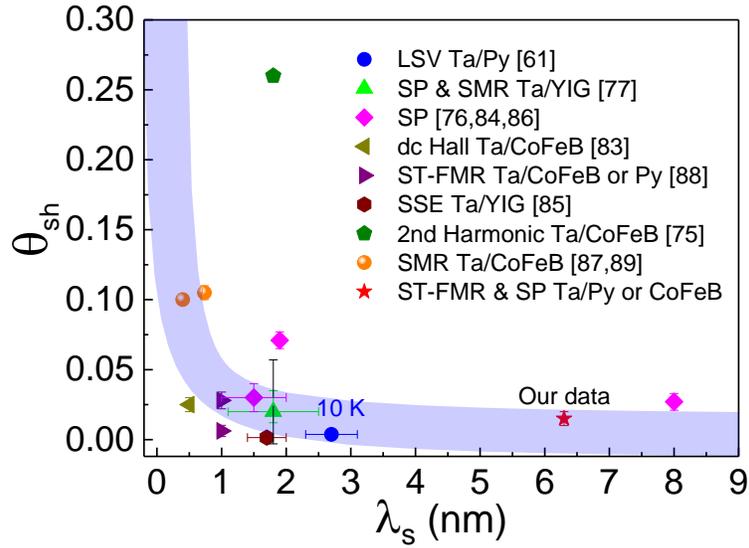

**Figure 11.** Summarized SOT efficiency $\theta_{sh}$ as a function of $\lambda_S$ in Ta from various reports. The LSV, SP, SMR, dc Hall, ST-FMR, SSE represent nonlocal measurement in lateral spin valves, spin pumping, spin Hall magnetoresistance, Hall measurements by using dc currents, spin torque ferromagnetic resonance, and spin seebeck effect method, respectively. The blue thick curve shows a correlation of $\theta_{sh}\lambda_S \sim 0.036$ nm. The $\theta_{sh}$ and $\lambda_S$ were obtained at room temperature, except for the one by the LSV method which was measured at 10 K as denoted in the figure.

### 5.3. Spin Hall effect in tungsten (W) and other alloys with heavy metal dopants

For the case of pure HMs, the largest SOT efficiency is ~0.3 observed in a *β*-phase W/CoFeB bilayer by ST-FMR measurements at room temperature [35]. This value is about two times larger than that in Ta, leading to efficient spintronic devices. Taking advantage of the highly efficient spin current generation, SOT driven magnetization switching has been demonstrated in a *β*-phase W based 3-terminal MTJ device [35]. The critical switching current density, $J_C$, is in the order of ~$10^6$ A/cm$^2$ in W/CoFeB/MgO [90], which is almost one order of magnitude smaller than that in Pt.

Figure 12 shows a summary of the SOT efficiencies in W, ranging from 0.0043 to 0.95, measured by different techniques from different research groups [84, 85, 87, 89-93]. However, the correlation between $\theta_{sh}$ and $\lambda_S$ is not as clear as the case of Pt as shown in figure 8(c). Similar to the case of Ta, the reason can be attributed to the multi-phases (such as $\alpha$-, $\beta$-phases) in various W thin films [35, 90].

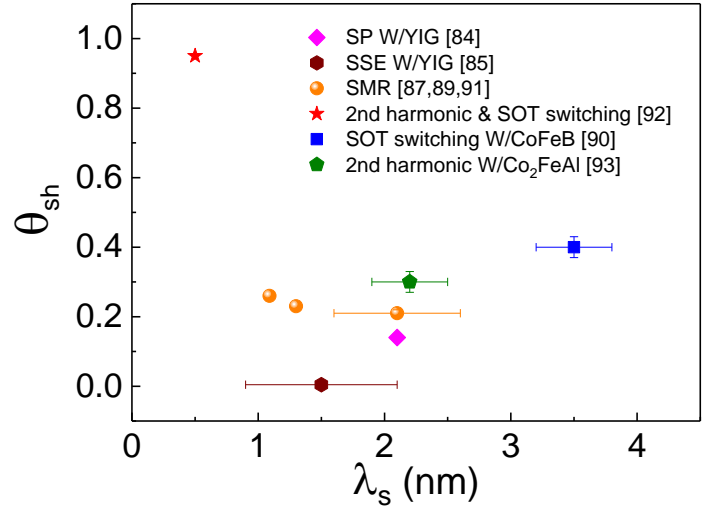

**Figure 12.** Summary of SOT efficiency $\theta_{sh}$ as a function of $\lambda_S$ in W from various reports. The SOT switching represents current induced magnetization switching via SOT. All data were obtained at room temperature.

Besides the pure HMs, alloys consisting of a light metal with heavy metal dopants have been also investigated [94-97]. Ramaswamy *et al.* [97] experimentally studied the effects on the SOT efficiency due to a systematic addition of Pt into the light metal Cu by ST-FMR measurements. As shown in figures 13(a) and (b), the SOT efficiency of $Cu_{1-x}Pt_x$ increases as the Pt concentration increases from 0 to 40%. It was found that only 28% Pt in $Cu_{1-x}Pt_x$ can give rise to a SOT efficiency close to that of Pt. Since Cu is the most common metallization element in the CMOS platform and $Cu_{1-x}Pt_x$ can have a large enough SOT efficiency, the $Cu_{1-x}Pt_x$-based alloy is easier to integrate the spintronic devices into an existing Si fabrication technology. In addition, from further analysis of the ST-FMR data, it is found that the skew scattering contribution is significant for lower Pt concentrations, while the side-jump contribution is significant for higher Pt concentrations. This result can be understood as due to the different scaling of the contributions of skew scattering and side-jump with respect to Pt concentrations. As Pt concentration increases, the longitudinal resistivity increases. In a simplistic picture, the skew-scattering contribution is independent of this increased resistivity due to Pt, whereas the side-jump contribution is proportional to this increased resistivity [98, 99]. Thus, for higher Pt concentrations, the increase in the resistivity is larger, leading to the domination of the side-jump contribution. Furthermore, this result can be also correlated with an earlier theoretical report [100] and with experiments in the context of anomalous Hall effect due to rare earth impurities in Gd [101].

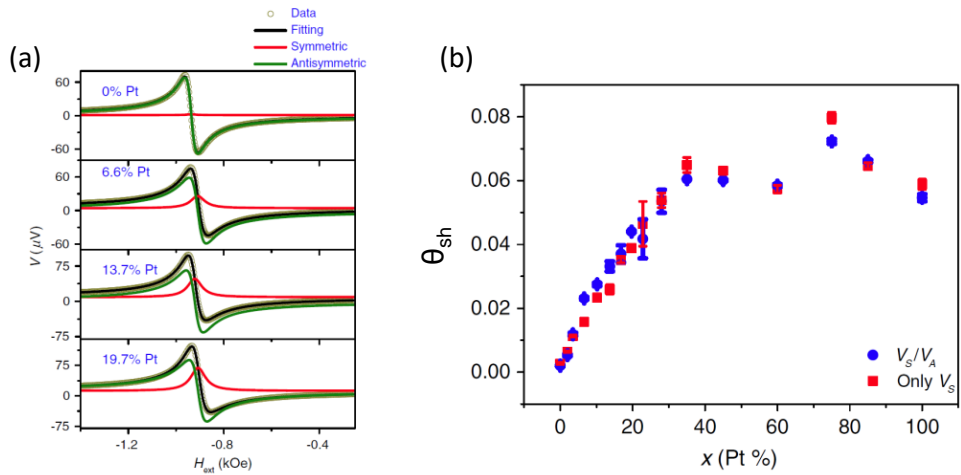

**Figure 13.** Fittings of ST-FMR signals from the $Cu_{1-x}Pt_x$ (6 nm)/Py (5 nm) bilayer for x = 0, 6.6, 13.7, and 19.7% in the negative $H_{ext}$ range for an applied microwave power of 16 dBm and a microwave frequency of 8 GHz. As the Pt concentration increases, the amplitude of the symmetric component increases (red curve). (b) SOT efficiency, $\theta_{sh}$, for different Pt concentrations extracted from ST-FMR fittings by using the $V_S/V_A$ ratio method (blue circles) and only-$V_S$ method (red squares), suggesting there is a negligible $\tau_{FL}$ at the interface between $Cu_{1-x}Pt_x$ and Py. Reprinted with permission from [97]. Copyright (2017) by the American Physical Society.

## 5.4. Role of interface transparency of HM and FM in the SOT efficiency

It is possible for the spin loss to occur at the interfaces when the spins transmit into the FM layer from the spin generation sources [70, 79], which can have an important role in the estimated

SOT efficiency. Using the ST-FMR technique, Zhang *et al.* [80] studied the transparency of Pt (6 nm)/Py (5.5 nm) and Pt (6 nm)/Co (5.2 nm), shown in figures 14(a) and (b), respectively. Using standard ST-FMR analysis method, the authors find that the SOT efficiency in Pt/Co is ~0.11, much larger than ~0.05 in Pt/Py. They ascribed the discrepancy of the SOT efficiency to the interface transparency, which is correlated to the effective spin-mixing conductance, $G_{eff}$ at the Pt and FM interface. They quantified the $G_{eff}$ to be $3.96 \times 10^{19}$ m$^{-2}$ (for Pt/Co) and $1.52 \times 10^{19}$ m$^{-2}$ (for Pt/Py), and the interface transparencies to the spin currents are ~0.65 (Pt/Co) and ~0.25 (Pt/Py), respectively. After taking into account the transparencies of these interfaces, they found that the intrinsic SOT efficiency in Pt has a much higher value of $0.17 \pm 0.02$ in Pt/Co and $0.20 \pm 0.03$ in Pt/Py bilayer. Further, this result also suggests that the spins easily transmit through the Pt/Co interface and thus give a large measured SOT efficiency, which might be due to better matching of the electronic bands between the Pt and Co layer compared to the Pt and Py layer.

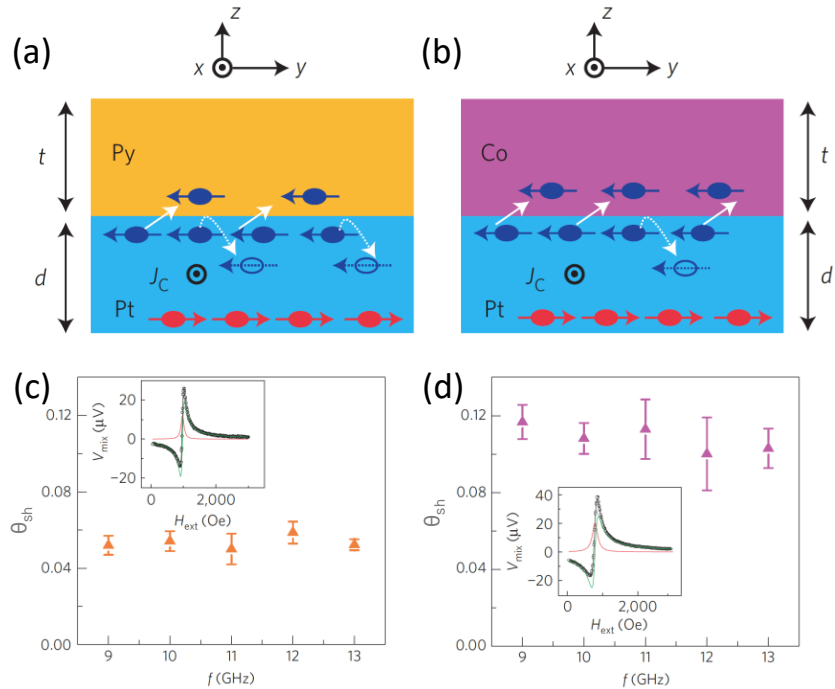

**Figure 14.** Illustration of the interface transparency for (a) Pt/Py and (b) Pt/Co: the red and blue arrows represent the up and down spin accumulation. The white arrows show the electrons diffusing across or being reflected at the interface. Frequency dependence of measured SOT efficiencies for (c) Pt (6 nm)/Py (5.5 nm) and (d) Pt (6 nm)/Co (5.2 nm). Insets: measured ST-FMR signals for Pt/Py and Pt/Co at 9 GHz, respectively. Reprinted by permission from Macmillan Publishers Ltd: Nature Physics [80], copyright 2015.

Subsequently, Pai *et al.* [81] also conducted a systematic study of the interface transparency in Pt/Co and Pt/CoFe bilayers. They found that the SOT efficiency can be modulated under different interface conditions and a much larger intrinsic SOT efficiency (~0.3) in Pt is obtained after considering the interface transparency between the Pt and FM layer. This value is much larger than the measured value of ~0.06 [25, 32], suggesting that there is a large spin memory loss at the interface. Therefore, it is understood that the measured SOT efficiency by ST-FMR technique represents a lower value because of the interface transparency. Therefore, modification or enhancement of the SOT efficiency via interface transparency engineering, such as inserting layers between the HM and FM [73, 74], manipulation of the film crystal structures [102] or using different combinations of HM and FM layers [80, 81], are some interesting and important on-going research activities.

## 5.5. Difference in the SOT efficiency between ST-FMR and magnetization switching technique

From ST-FMR measurements, $\tau_{DL}$ and $\tau_{FL}$ can be separately extracted due to the 90° phase difference in the magnetization dynamics (see Section 2.3) [33, 40]. Subsequently, $\tau_{DL}$ can be used to evaluate the in-plane SOT efficiency $\theta_{sh}$. However, for the case of current driven magnetization switching using SOT, it has been reported that both $\tau_{DL}$ and $\tau_{FL}$ contribute to the magnetization switching and the switching current density $J_C$ might be significantly reduced if there is an additional $\tau_{FL}$ [103-106], which can lower the energy barrier of magnetization switching. Therefore, the values of the SOT efficiency derived from SOT induced magnetization switching measurements can be greater than that from ST-FMR measurements. In addition, during the ST-FMR measurements, the magnetization is uniformly aligned in the direction of an applied large external magnetic field. Hence, it is a coherent magnetization precession process. However, the current induced magnetization switching proceeds through an thermally excited incoherent magnetization switching process [107, 108] (i.e. domain wall motion with a much lower energy barrier), which can also lead to a smaller $J_C$ and a larger $\theta_{sh}$ from magnetization switching measurements compared to the ST-FMR measurements. Therefore, these aspects should be taken into account for the SOT efficiency evaluation or comparison between ST-FMR and switching measurements.

## 6. SOT in topological insulators (TIs)

Topological insulators (TIs) are quantum materials that have a band gap just like a normal insulator but have topologically protected conducting edge states or surface states resulting from the inverted conduction and valence bands due to strong SOC [109-112]. TI materials started from two-dimensional (2D) system, called 2D TI. The 2D TI was predicted [113] and experimentally observed in HgTe/CdTe quantum wells structures in 2007 [109]. As shown in figure 15(a), two conductive edge states are present in the edge of 2D TIs, each of which contributes one quantum of conductance $e^2/h$, where $h$ is the Plank's constant. In the edge states, the spin polarization depends on the electron momentum.

A year later, in 2008, the first 3D topological insulator $Bi_{1-x}Sb_x$ was experimentally discovered [114]. Since then, more 3D TIs such as $Bi_2Se_3$, $Bi_2Te_3$, and $Sb_2Te_3$, having a larger bandgap and single Dirac cone at the $\Gamma$ point, have been predicted and identified mainly by angle resolved photoemission spectroscopy (ARPES) experiments [110-112]. The 3D TIs possess spin-momentum-locked topological surface states (TSS) due to time reversal symmetry protection [115-117] and it can be described by the Dirac Hamiltonian $H_k = v_F(\hat{z} \times \hat{\sigma}) \cdot k$, where $k$ is electron momentum, $\hat{z}$ is the unit vector perpendicular to the TI films and $v_F$ is the Fermi velocity. This dispersion relationship reveals that on the TSS, the electron momentum and the spin polarization directions are strongly locked as shown in figure 15(b). As depicted in figure 15(c), in real space, as charge currents flow on TSS, all the electron spins are expected to be fully polarized in the orthogonal direction to the electron moving direction due to the topologically protection [118]. Therefore, a very efficient spin current generation and thus a giant SOT efficiency are expected in TIs due to TSS. For further details into TIs and related physics phenomena, the readers can refer to detailed reviews on TIs [116, 117]. In the following part, we will review research works on 3D TIs in the context of SOT efficiency determination.

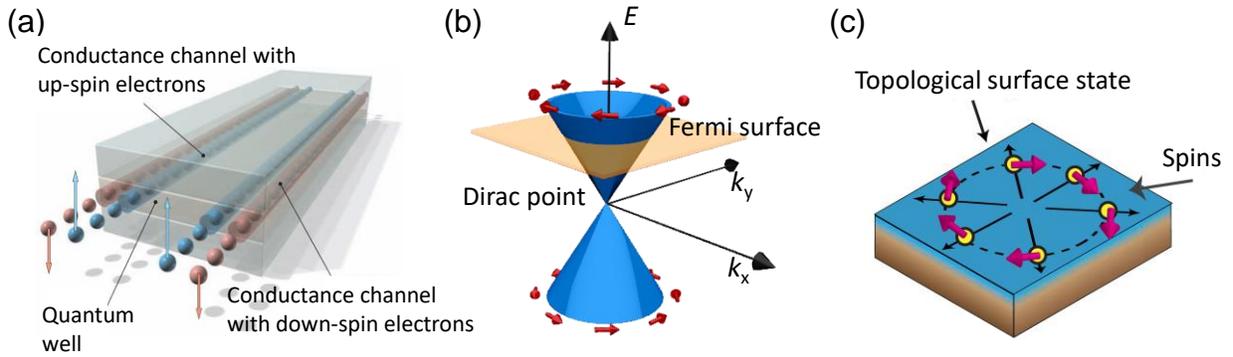

**Figure 15.** (a) Schematic of the spin momentum locked edge states in a quantum spin Hall insulator (i.e. 2D TI). From [109]. Adapted with permission from AAAS. (b) Schematic of Dirac cone of the TSS, illustrating the Dirac point (DP), Fermi level ($E_F$) and spin momentum locking in momentum space. Reprinted with permission from [33]. Copyright (2015) by the American Physical Society. (b) Illustration of the electron flow and spin polarizations on the surface of TI for opposite applied charge currents. Adapted with permission from [118]. Copyright (2014) by the American Physical Society.

## 6.1. SOT efficiency in TIs

Research works have been conducted to demonstrate the TSS and quantify the SOT efficiency in TI/FM heterostructures [33, 34, 40, 118-126]. In 2014, Mellnik *et al.* [40] studied the $Bi_2Se_3$ (8 nm)/Py (16 nm) bilayers by ST-FMR measurements at room temperature. Figures 16(a) and (b) show the schematic of the bilayer structure and the representative ST-FMR signal. Using the only-$V_S$ analysis method, the in-plane SOT efficiency, $\theta_{sh}$, was determined to be ~2.0–3.5 at room temperature, which is about one to two orders of magnitude larger than that in HMs reported previously. In addition, the out-of-plane SOT efficiency, $\theta_\perp$, arising from $\tau_{FL}$ has the similar order of amplitude as $\theta_{sh}$. The spin configuration is consistent with that expected in the TSS of TIs [115-117]. However, the exact mechanisms for this large SOT efficiency observed in $Bi_2Se_3$ was not clear at that time.

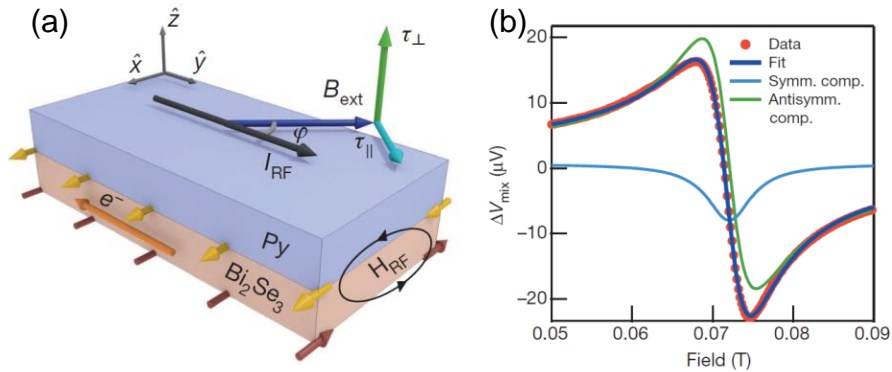

**Figure 16.** (a) Schematic diagram of $Bi_2Se_3$ (8 nm)/Py (16 nm) bilayer structure. The yellow and red arrows denote spin moment directions. (b) Measured ST-FMR signal at room temperature with $f = 8$ GHz and $\varphi = 45°$. A fixed microwave power of 5 dBm is absorbed by the device, corresponding to $I_{RF} = 7.7 \pm 1.1$ mA. The lines are the fits showing the symmetric and antisymmetric components. Reprinted by permission from Macmillan Publishers Ltd: Nature [40], copyright 2014.

Subsequently, Wang *et al.* [33] verified that the observed SOTs originate from the TSS in $Bi_2Se_3$ by temperature-dependent ST-FMR measurements on a $Bi_2Se_3$ (20 nm)/CoFeB (5 nm) heterostructure. By taking the only-$V_S$ and -$V_A$ analysis method, the authors determined both $\theta_{sh}$ and $\theta_\perp$ as a function of temperature from three representative ST-FMR devices. As shown in figure 17(a), the $\theta_{sh}$ was found to increase steeply and nonlinearly, especially below 50 K, reaching the value of ~ 0.42, which is almost 10 times larger than that at 300 K. Moreover, a difference in the $\theta_{sh}$ values between two ST-FMR analysis methods indicates a noticeable $\tau_{FL}$ (i.e. $\theta_\perp$), which is plotted against temperature in figure 17(b). As it is known that the $\theta_{sh}$ arising from the bulk spin Hall mechanism normally shows a very weak temperature dependence [32, 55, 96], and the $\theta_\perp$ arising from the field-like torque or interfacial Rashba effect usually shows a smaller value at low temperature in HMs or semiconductors [75, 127, 128]. In addition, the hexagonal warping effect in the TSS can account for the out-of-plane spin polarizations and thus affect the $\theta_\perp$ [129-132]. Therefore, the significant increase of $\theta_{sh}$ and $\theta_\perp$ in the Bi$_2$Se$_3$/CoFeB bilayer as temperature decreases is attributed to the TSS in Bi$_2$Se$_3$. These results suggest that the TSS with spin momentum locking can play an important role in the spin current generation. This work suggests that the bulk states (BS) in Bi$_2$Se$_3$ and the 2DEG on top of Bi$_2$Se$_3$ can lead to a contamination to the SOT effects from TSS. In Section 6.3, we present the role of BS, 2DEG and TSS of Bi$_2$Se$_3$ in the SOT efficiency by ST-FMR measurements.

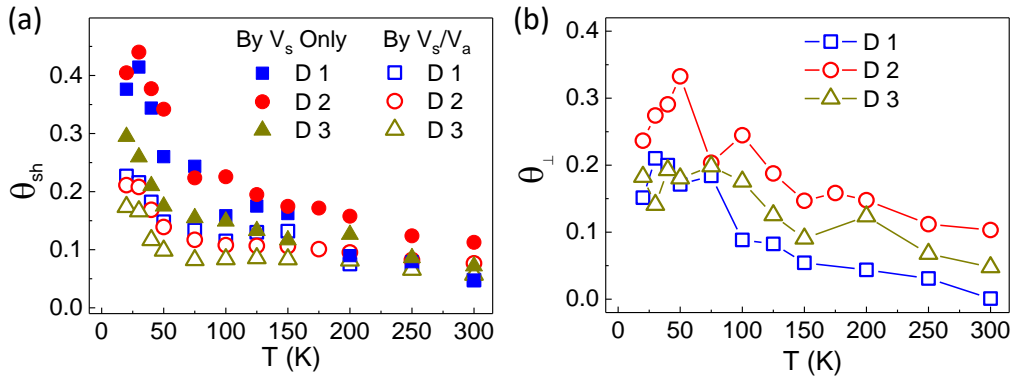

**Figure 17.** Temperature dependence of (a) in-plane SOT efficiency, $\theta_{sh}$, and (b) out-of-plane SOT efficiency, $\theta_\perp$, in Bi$_2$Se$_3$ (20 nm)/CoFeB (5 nm) for three devices: D1, D2, and D3. The SOT

efficiencies are analyzed by two different methods: by only-$V_S$ and by $V_S/V_A$ ratio. Reprinted with permission from [33]. Copyright (2015) by the American Physical Society.

The Fermi level dependent SOT efficiency was recently studied in a $(Bi_{1-x}Sb_x)_2Te_3$ (BST, 8 nm)/Cu (8 nm)/Py (10 nm) [120]. As shown in figure 18(a), the Fermi level position relative to the Dirac point was controlled by using different Sb compositions. From ST-FMR measurements on each device, it was observed that the *interface* charge-to-spin conversion efficiency, $q_{ICS}$ due to the TSS exhibits a large and nearly constant value for $0 < x < 0.7$ (see figure 18(b)), suggesting the highly efficient charge and spin interconversion in the TSS of TI materials. However, the $q_{ICS}$ was found remarkably reduced as the Fermi level traverses through the Dirac point, which is possibly due to the inhomogeneity of $k_F$ and/or instability of the helical spin structure near the Dirac point.

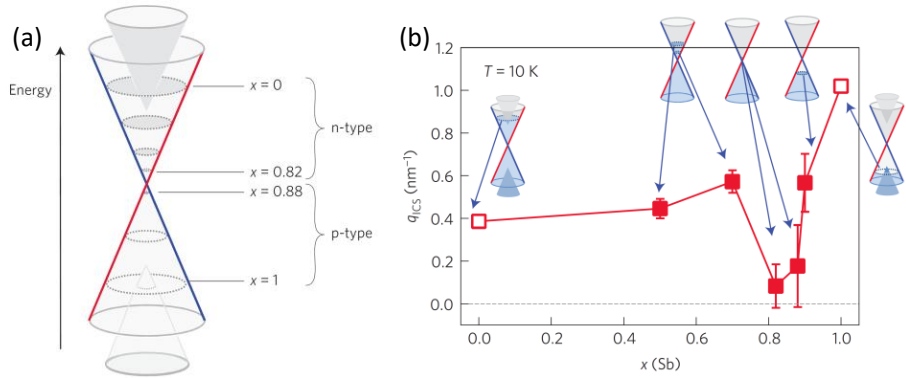

**Figure 18.** (a) Schematic of the energy dispersion and Fermi level position in $(Bi_{1-x}Sb_x)_2Te_3$ with different Sb composition. (b) Interface charge-to-spin conversion efficiency, $q_{ICS}$ as a function of Sb composition. The inset shows the band structure and Fermi level position for each Sb composition. Bulk insulating BST with $0.50 \leq x \leq 0.90$ should have only surface transport. $Bi_2Te_3$ ($x = 0$) and $Bi_2Sb_3$ ($x = 1$) have both bulk and surface conduction paths. Reprinted by permission from Macmillan Publishers Ltd: Nature Physics [120], copyright 2016.

### 6.2. Anisotropic SOT (efficiency) in $Bi_2Se_3$ MBE films

$Bi_2Se_3$ has a three-fold symmetry when viewed from the top (along the *z*-axis) shown in figures 19(a) and (b). Therefore, the molecular beam epitaxy (MBE) grown $Bi_2Se_3$ films naturally have triangular islands as captured by atomic force microscope in figure 19(c). It is found that the orientation of the triangular islands prefers to a certain axis with twin structures due to the six-fold symmetric sapphire substrate. Although the anisotropy of electron transport has been observed in $Bi_2Se_3$ films previously [133], the anisotropy of SOT in $Bi_2Se_3$ has not been reported.

We have performed ST-FMR to study the anisotropic SOT characteristics in high quality MBE grown $Bi_2Se_3$ films shown in figure 19(c). $Bi_2Se_3$ Hall bars with different angle $\varphi$, the angle between the channel current ($I$) direction with respect to the $x$-axis denoted in figure 19(c), were fabricated in the same sample. Similarly, using the same $Bi_2Se_3$ wafer, ST-FMR devices of $Bi_2Se_3$ (10 QL)/CoFeB (7 nm) devices with different angle $\varphi$ were also fabricated for the ST-FMR measurements. As shown in figures 19(d-e), both the resistivity ($\rho_{BiSe}$) and sheet carrier concentration ($n_{2D}$) show angular dependent behaviors with the oscillation period of 60°. Furthermore, the in-plane SOT efficiency, $\theta_{sh}$, in figure 19(f) also shows an oscillation behavior as the current $I$ flows along different crystal axes (i.e. different angle $\varphi$). The oscillation period is also 60°, in which $\theta_{sh}$ has a low value at $\varphi = 0°$ but a high value at $\varphi = 30°$, and returns to a low value at $\varphi = 60°$. This is consistent with the six-fold symmetry by considering the $Bi_2Se_3$ twin structures. Thus, this oscillation of the SOT efficiency is attributed to the hexagonal warping in the Fermi surface of the $Bi_2Se_3$ films [129, 132, 134]. An out-of-plane spin polarization in the TSS has been experimentally observed in $Bi_2Se_3$ [130, 131] due to the hexagonal warping effect, which shows a finite amplitude along the Γ-K direction but almost zero along the Γ-M direction in momentum space, leading to an oscillation period of 60° in the out-of-plane spin polarization. If we assume that the total spin polarization is unity in TSS, the in-plane spin polarization should also show the same oscillation period of 60°, which is in line with our observation.

Note that since the width of the current channel of the Hall bars and ST-FMR devices is 20

μm, which is much larger than the triangle size of ~100–200 nm in figure 19(c), our data only show the averaged transport results. In addition, there exists nonuniform $Bi_2Se_3$ microstructures and devices with different $\varphi$ are located different places on a sample. Consequently, a fluctuation of the high (or low) values at each angle $\varphi$ can be observed in figures 19(d-f). Note that the similar anisotropy behavior is expected in the range of $\varphi = 180 - 360°$ due to the crystal symmetry. Rather than rotating the crystal direction in different devices, the angular dependent ST-FMR measurements in one device can be combined to characterize the anisotropy of SOTs [135], which will be discussed in Section 7.2. Therefore, the ST-FMR is a suitable dynamic measurement method to study the anisotropy of SOTs.

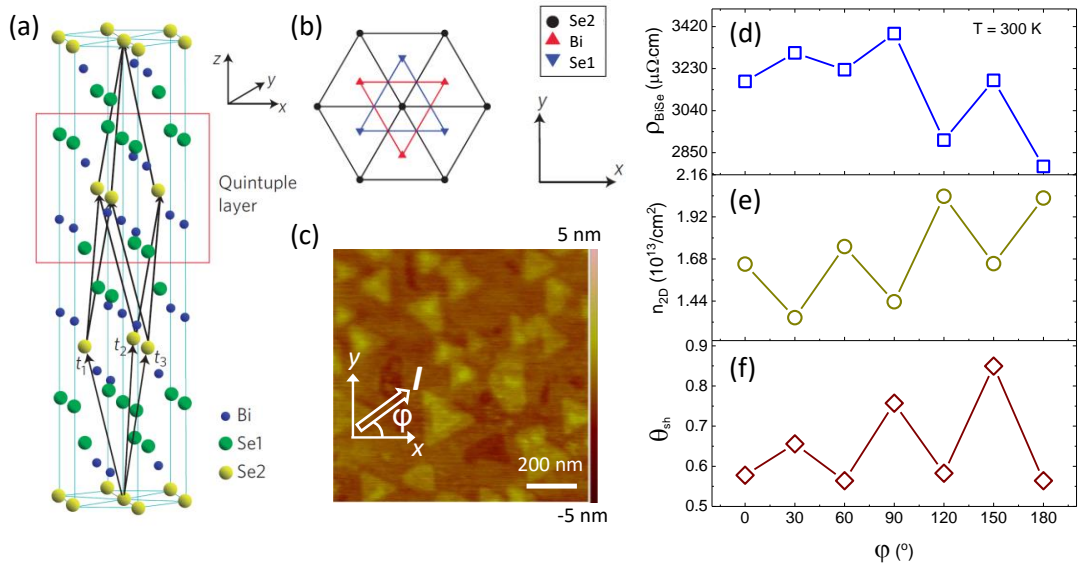

**Figure 19.** (a) Crystal structure of $Bi_2Se_3$ with three primitive lattice vectors denoted as $t_{1;2;3}$. A quintuple layer (1 QL ≈ 1 nm) with Se1–Bi1–Se2–Bi1'–Se1' is indicated by the red square. Adapted by permission from Macmillan Publishers Ltd: Nature Physics [111], copyright 2009. (b) Top view along the $z$-direction. The triangle lattice in 1 QL has three different positions, denoted as Se2, Bi and Se1. Adapted by permission from Macmillan Publishers Ltd: Nature Physics [111], copyright 2009. (c) Atomic force microscope image of a 10-QL $Bi_2Se_3$ film with a roughness of ~0.5 nm. Triangular islands are observed with a clear terrace step of 1 QL, indicating a high quality $Bi_2Se_3$ film. (d-f) The $Bi_2Se_3$ resistivity, sheet carrier concentration and in-plane SOT efficiency

measured with charge currents flowing along different crystalline directions. The angle $\varphi$ is denoted in (c).

### 6.3. TSS dominated SOT

As mentioned earlier, the BS, 2DEG, and TSS usually coexist in TIs such as $Bi_2Se_3$. Therefore, it is important to understand the role of each channel on the SOT efficiency in TIs, so that one can maximize the SOT efficiency to realize highly efficient SOT driven magnetization switching. Figure 20 shows a summary of the reported SOT efficiencies in various TIs with different thicknesses. The SOT efficiency increases significantly as TI films become thinner even though SOT efficiencies are evaluated using different techniques and in different TIs. Recently, the SOT efficiency of MBE grown $Bi_2Se_3$ films with various thicknesses in the range of 5 to 20 QL has been quantified by ST-FMR measurements [34]. The optimum thickness range of $Bi_2Se_3$ has been identified to be 5–8 QL to maximize the SOT effect in $Bi_2Se_3$ devices. Figure 21(a) shows the schematic of the film stack used for the ST-FMR measurements. First, it was found that the resistivity, $\rho_{BiSe}$ rapidly increases when the $Bi_2Se_3$ thickness ($t_{BiSe}$) is less than 10 QL as shown in figure 21(b). On the other hand, the sheet carrier concentration ($n_{2D}$) shows an opposite trend, decreasing substantially below 10 QL as shown in figure 21(c). The carrier concentrations from BS ($n_{2D-Bulk}$), 2DEG ($n_{2DEG}$) and TSS ($n_{TSS}$) were further separated and it was found that in the region of 5–8 QL, the value of $n_{TSS}$ is larger compared to $n_{2DEG}$ and $n_{2D-Bulk}$, indicating a TSS dominated electrical transport in thin $Bi_2Se_3$ films.

Figure 21(d) presents $\theta_{sh}$ as a function of $t_{BiSe}$ at room temperature by ST-FMR measurements on $Bi_2Se_3$ ($t_{BiSe}$)/CoFeB (7 nm) bilayers. As $t_{BiSe}$ is in the range of 5–8 QL, the $Bi_2Se_3$ films exhibit a giant $\theta_{sh}$ of ~1–1.75 at room temperature, which further corroborates the TSS dominated transport in thin $Bi_2Se_3$ film region. In addition, the *interface* SOT efficiency from TSS ($\lambda_{TSS}$) using an interface charge current density $J_{C\text{-}TSS}$ (A cm$^{-1}$) in TSS were estimated for each $t_{BiSe}$ as shown by squares in figure 21(e). Furthermore, after subtracting the opposite 2DEG contribution, the intrinsic $\lambda_{TSS}$ was calculated as 0.8 nm$^{-1}$ for 7, 8 and 10 QL Bi$_2$Se$_3$, as shown by circles, which is similar to recently reported interface SOT efficiency values in (Bi$_{1-x}$Sb$_x$)$_2$Te$_3$ [120].

These results suggest that the BS and 2DEG dilute the TSS contribution and weaken the SOT efficiency in Bi$_2$Se$_3$, and thus by using moderately thin TI films (5-8 QL) one can obtain the TSS dominated SOTs for device applications. In addition, the TSS dominated transport has also been reported recently [123] by changing the Bi$_2$Se$_3$ film thickness with a spin pumping technique, which is a reverse process (spin-charge conversion) of ST-FMR.

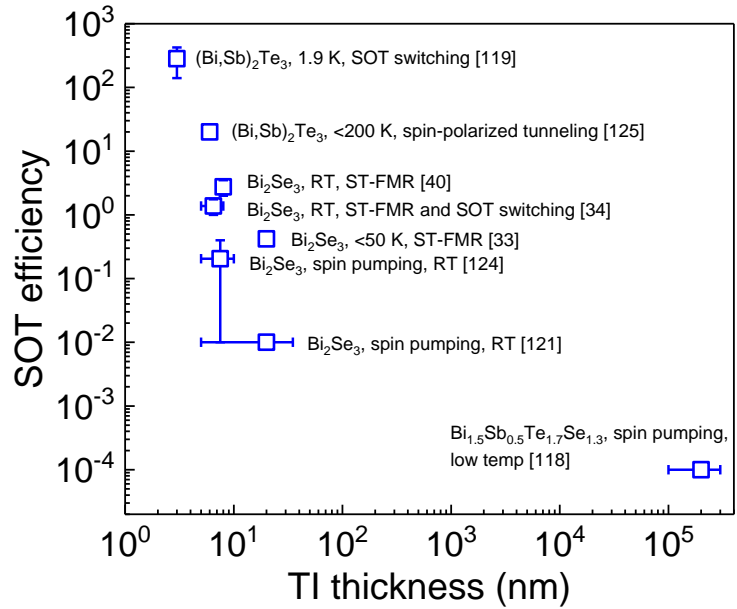

**Figure 20.** Summary of the SOT efficiencies measured in different TIs with different thicknesses. The materials with corresponding measurement temperature, measurement method and reference are indicated. SOT switching represents the SOT driven magnetization switching.

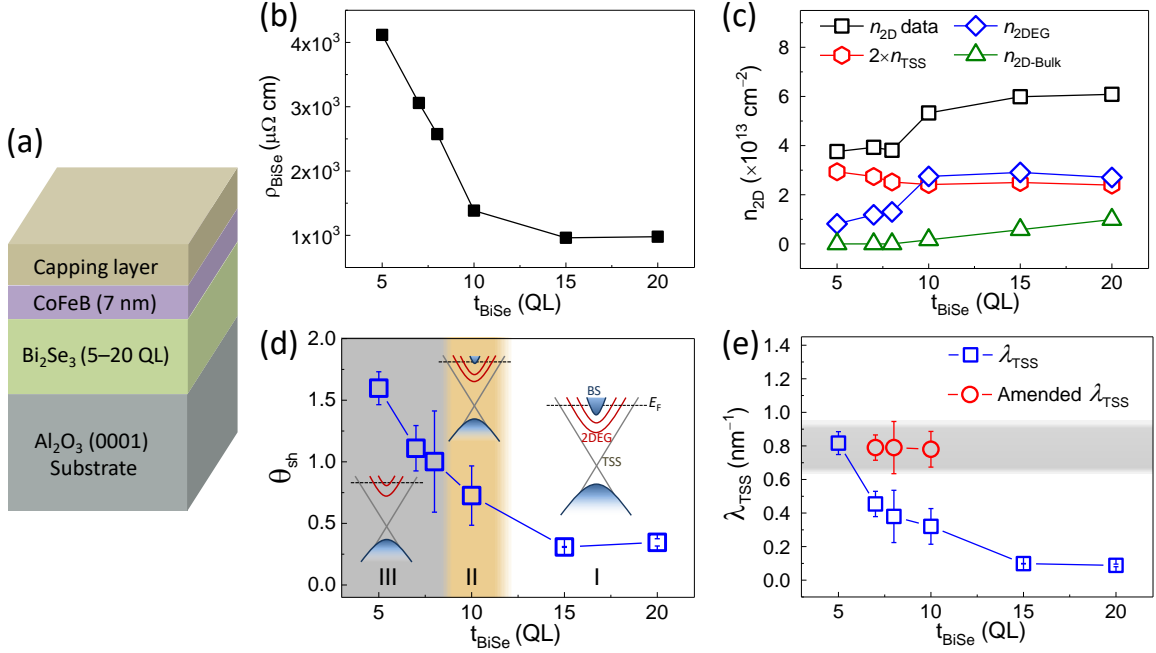

**Figure 21.** (a) Schematic of the film stack for ST-FMR devices. (b) $Bi_2Se_3$ thickness dependent resistivity, $\rho_{BiSe}$, in $Bi_2Se_3$ at room temperature. (c) Estimation of sheet carrier concentration of TSS, 2DEG and bulk channels. (d) $\theta_{sh}$ as a function of $Bi_2Se_3$ thickness at room temperature. Each $\theta_{sh}$ represents the averaged value from three devices. Region I, II and III denoted by different colours represent the charge-to-spin conversion dominated by different mechanisms. The inset shows the schematic of the band structure for each region. (e) Interface SOT efficiency, $\lambda_{TSS}$ (squares), as a function of $t_{BiSe}$ at room temperature. Amended interface SOT efficiencies from TSS after excluding the opposite 2DEG contribution for 7, 8 and 10 QL $Bi_2Se_3$ with circles. Adapted from [34].

By taking advantage of this giant $\theta_{sh}$, SOT driven magnetization switching in the $Bi_2Se_3$ (8 QL)/Py (6 nm) heterostructures was successfully achieved at room temperature without any external magnetic field [34]. This work makes a substantial improvement in the working temperature of TI based magnetization switching scheme from 1.9 K [119] to 300 K, demonstrating that TI can be an excellent spin generator at room temperature. The critical switching current density $J_C$ in $Bi_2Se_3$ required for the magnetization switching is extremely low ($\sim 6\times 10^5$ A/cm$^2$), which is almost two orders of magnitude smaller than that in HMs, such as Pt and Ta [20-22]. The much lower $J_C$ for magnetization switching using TIs is promising to address the outstanding scalability and power consumption issue in modern magnetic devices. Furthermore, the magnetization switching scheme demonstrated in Ref. [34] does not require an assistive magnetic field, which makes the TI/FM material systems easy to integrate into the well-established industrial technology for magnetic devices. In addition, the room temperature SOT driven magnetization switching has been also observed recently in the $Bi_2Se_3$/ferrimagnetic CoTb heterostructures [136] and in the sputtered $Bi_xSe_{(1-x)}$/Ta/CoFeB/Gd/CoFeB heterostructures [137].

## 7. SOT beyond HMs and TIs

Beyond HMs and TIs, other novel material systems with strong SOC are emerging, which also have potential in the future spintronic applications. The following sections discuss some representative examples, in which the ST-FMR technique was used to evaluate the SOTs (or SOT efficiency), to get insight into the underlying physics of SOTs in these emerging systems.

### 7.1. Interface between two nonmagnetic materials

The Rashba SOC, which was first discussed in the 2DEG systems with spin degeneration lifting [138], can emerge at the interfaces in a variety of material systems nowadays [139]. The Rashba SOC at an interface is explained as follows. For an interface with inversion symmetry breaking in the direction perpendicular to the interface, an electric field $E$ is produced perpendicular to the interface. An electron moving within this interface with the electric field will experience an effective magnetic field $H$ due to the relativistic corrections, which causes a momentum dependent spin splitting in the bands, analogous to the Zeeman splitting. In the scheme of Rashba SOC, the interaction between the spin polarization $\hat{\sigma}$ and electron momentum $k$ can be expressed by the Hamiltonian $H_R = \alpha_R (k \times \hat{z}) \cdot \hat{\sigma}$, where $\alpha_R$ is the Rashba coefficient, $\hat{z}$ is the unit vector perpendicular to the interface. Figures 22(a) and (b) schematically show the dispersion curves with spin splitting due to the Rashba SOC at the interface and the corresponding Fermi contours with spin polarization configuration. It is revealed that the spins polarizations are transverse to the electron moving direction. Since the spins have different chiralities in two contours, they compensate each other and give rise to the net spin polarizations, which can be measured in the experiments.

The Rashba SOC at the interface between the HM and FM layer, such as Pt/Co, has been studied previously [20, 72]. Recently, researchers have found that when one nonmagnetic material, such as Bi, Pb or Sb, contacts with other nonmagnetic material, such as Ag, their interface can also exhibit strong interface SOC [140-147]. In 2015, the charge-to-spin conversion induced by the Rashba-Edelstein effect was directly observed at the interface of Bi/Ag [146]. Using a spin-polarized positron beam, they found an opposite surface spin polarization between Bi/Ag/Al$_2$O$_3$ and Ag/Bi/Al$_2$O$_3$ samples. Subsequently, Jungfleisch *et al.* [147] reported the Rashba-Edelstein-induced-spin driven ST-FMR in Bi/Ag/Py heterostructures. They evaluated the SOT efficiency of the Bi/Ag interface to be ~0.18 using the $V_S/V_A$ ratio method.

Another recent report shows a very large SOT efficiency at the interface between 10 QL Bi$_2$Se$_3$ TI materials and nonmagnetic material Ag thin layer at room temperature by ST-FMR measurements [148]. As shown in figure 22(c), from first-principle calculations, it was found that there is a pair of large Rashba splitting bands emerging at the interface between Bi$_2$Se$_3$ and Ag. Moreover, the Rashba bands are located outside the TSS linear bands of the Bi$_2$Se$_3$ layer, and the Rashba bands have the same net spin polarization direction as the TSS of Bi$_2$Se$_3$. As shown in figure 22(d), due to the large interface Rashba SOC at the newly formed Bi$_2$Se$_3$/Ag interface, the measured SOT efficiency shows a significant enhancement as the Ag insertion layer thickness increases to ~2 nm and reaches a value of 0.5 for 5 nm Ag. The SOT efficiency in Bi$_2$Se$_3$/Ag (5

nm)/CoFeB is ~3 times higher than that from sole TSS in Bi$_2$Se$_3$/CoFeB at room temperature. The extracted $α_R$ at the Bi$_2$Se$_3$/Ag interface has a significant large value about 2.83−3.83 eV Å and this value is similar to that of Bi/Ag previous reported [141]. Furthermore, Rashba-induced magnetization switching in Bi$_2$Se$_3$/Ag/Py with a low current density of $5.8 \times 10^5$ A/cm$^2$ has been demonstrated. As indicated in Section 6.3, the SOT efficiency in Bi$_2$Se$_3$/Ag can be further enhanced by decreasing the Bi$_2$Se$_3$ thickness to eliminate the bulk contamination and energy consumption will be further decrease for SOT driven magnetization switching. These reports indicate that the nonmagnetic bilayer interface can be designed to provide a large Rashba spin splitting and thus giant SOT efficiency, which has the potential to be used as efficient spin current sources for future spintronic devices.

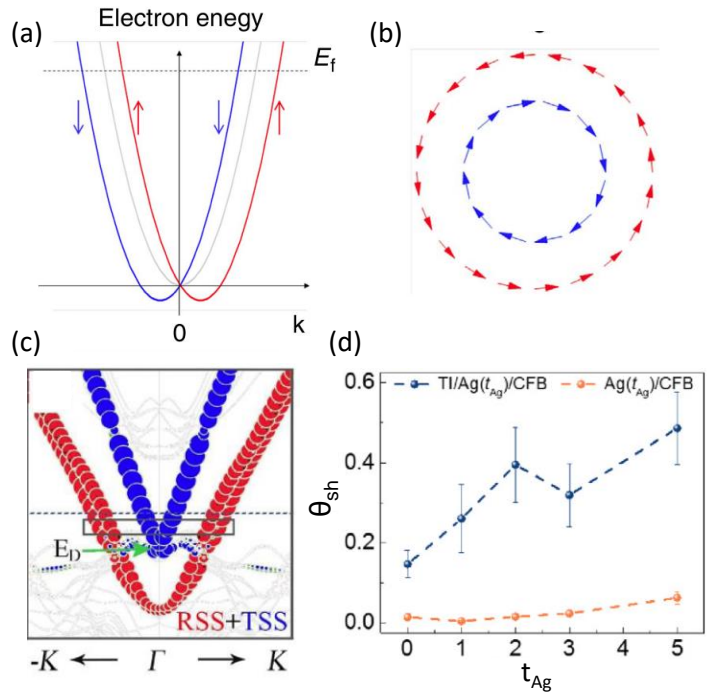

**Figure 22.** (a) Typical spin-split dispersion curves of a Rashba 2DEG. (b) Fermi contours with spin polarization denoted by the blue and red arrows. Reprinted by permission from Macmillan Publishers Ltd: Nature Communications [143], copyright 2013. (c) Spin-orbit-torque efficiency obtained for both Bi$_2$Se$_3$ (10 QL)/Ag ($t_{Ag}$)/CoFeB (7 nm) and Ag ($t_{Ag}$)/CoFeB (7 nm) ST-FMR

devices. (d) The band structure of Bi$_2$Se$_3$ (6QL)/Ag (0.95 nm) from first principle calculations. Reprinted with permission from [148]. Copyright (2018) by the American Physical Society.

## 7.2. Transition-metal dichalcogenides (TMDs)

In the recent years, 2D transition-metal dichalcogenide (TMDs) materials are attracting interests due to the novel physics and potential applications, which were reviewed in Ref. [149]. Since the TMDs display 2D characteristics, the thickness of 2D-TMDs can be reduced as thin as a monolayer. Therefore, TMDs provide an interesting platform to observe exotic interface related phenomena. Since 2016, the 2D-TMDs have been studied as spin current sources for the SOT devices [135, 150-153]. Spin current induced ferromagnetic resonance has been observed in monolayer MoS$_2$/Py structures [151], where the Py layer was deposited by either magnetron sputtering or e-beam evaporation. It was observed that both $\tau_{DL}$ and $\tau_{FL}$ in the monolayer MoS$_2$/Py system were present, with the ratio of $|\tau_{FL}/\tau_{DL}| = 0.19$. The observed SOTs were attributed to the interface between the MoS$_2$ and Py layer. The ratio of $|\tau_{FL}/\tau_{DL}|$ in MoS$_2$/Py is much smaller than that recently reported in the Bi$_2$Se$_3$/Py (or CoFeB) systems [33, 40], whose value is in the range of 1.4–2. This implies that the damping-like torque $\tau_{DL}$ is much larger than $\tau_{FL}$ in MoS$_2$/Py system. However, the real SOT efficiency of MoS$_2$/Py bilayer was not evaluated in Ref. [151] and the real SOT efficiency should be estimated by the method only-$V_S$ (or $\tau_{DL}$) described in Section 4.2 for proper comparison with other material systems.

Semimetal WTe$_2$ is also a layered TMD, which has strong SOC [154, 155] and has been reported to possess exotic topological properties [156]. Compared to other TMDs, such as MoS$_2$, an interesting property of single crystal WTe$_2$ is the crystal symmetry. As shown in figure 23(a), the mirror symmetry exists only relative to the *bc* plane in WTe$_2$, and there is no mirror symmetry in the *ac* plane. Therefore, there is no 180° rotation about the *c*-axis (perpendicular to the sample plane). Consequently, due to the lack of two-fold rotational symmetry about the *c*-axis, there is a possible existence of the out-of-plane damping-like SOT ($\tau_B$) in WTe$_2$. Recently, MacNeill *et al.* [135] have verified the presence of $\tau_B$ in WTe$_2$/Py bilayer systems by ST-FMR measurements at room temperature shown in figures 23(b) and (c). It was found that when the charge current flows along the *b*-axis with mirror symmetry in a WTe$_2$ (5.5 nm)/Py (6 nm) device, both the symmetric ($V_S$) and antisymmetric ($V_A$) ST-FMR components at different in-plane magnetic field angle $\varphi$ follow the behaviour of $\cos\varphi \times \sin(2\varphi)$, which is similar to the case of the traditional HM/FM bilayers. However, as the charge current flows along the *a*-axis where mirror symmetry is broken, $V_A$ at different $\varphi$ exhibits a novel behaviour of $\cos\varphi \times \sin(2\varphi) + \sin(2\varphi)$, indicating the presence of $\tau_B$ in WTe$_2$/Py as shown in figure 23(d). Figure 23(e) shows the SOTs in WTe$_2$ with different thicknesses ranging from monolayer to ~16 nm [153]. The magnitude of $\tau_B$ shows a very weak WTe$_2$ thickness dependence, suggesting that it is of the interface origin.

Most of the studied interface systems so far only have broken inversion symmetry along their vertical structure, which leads to an in-plane damping-like torque or out-of-plane field-like torque due to Rashba-Edelstein effect at the interface. On the contrary, the results in Ref. [135] demonstrate the in-plane rotational symmetry breaking induced out-of-plane damping-like torque in WTe$_2$, which is a novel interface related SOT phenomenon and provides a platform for field-free SOT driven magnetization switching of FM with perpendicular magnetic anisotropy (PMA).

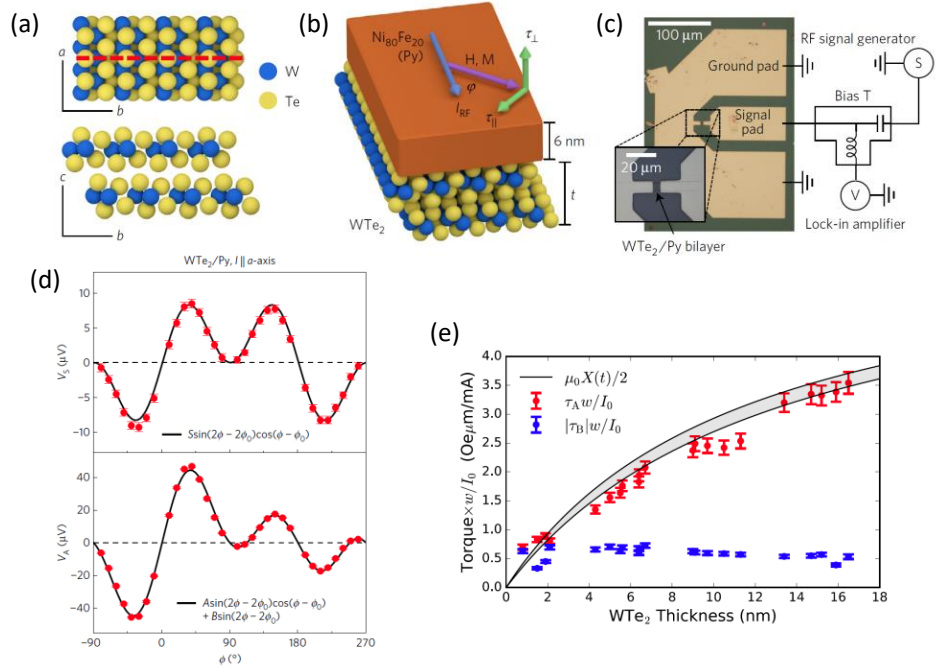

**Figure 23.** (a) Crystal structure near the surface of $WTe_2$. (b) Schematic of the $WTe_2$/Py sample. (c) The device and the circuit used for ST-FMR measurements. (d) Symmetric and antisymmetric ST-FMR components ($V_S$ and $V_A$) for a $WTe_2$ (5.5 nm)/Py (6 nm) device as a function of in-plane magnetic-field angle $\varphi$ denoted in (b). The charge current is applied parallel to the $a$-axis. The microwave frequency is 9 GHz and the applied microwave power is 5 dBm. Adapted by permission from Macmillan Publishers Ltd: Nature Physics [135], copyright 2016. (e) Out-of-plane damping-like torque ($\tau_B$, blue circles) and out-of-plane field-like torque ($\tau_A$, red circles) per unit interface charge current density ($I_0/w$) as a function of $WTe_2$ thickness, where $I_0$ and $w$ are charge current and channel width, respectively. Reprinted with permission from [153]. Copyright (2017) by the American Physical Society.

### 7.3. Antiferromagnetic (AFM) materials

Antiferromagnetic materials have zero net magnetization due to the antiparallel alignment of magnetic moments on the adjacent individual atoms. Therefore, the antiferromagnets are difficult to control by external magnetic fields. However, an injected charge current induced internal field can perturb the spin structures and thus magnetic ordering in AFM, as recently reported in different single antiferromagnetic (AFM) materials [157-159]. On the other hand, the devices with AFM/FM heterostructures have also been utilized to realize the external-magnetic-field-free SOT driven magnetization switching due to the presence of exchange bias [160-163], which breaks the magnetization up-down equivalence in the systems with PMA. Furthermore, a single AFM material itself can also work as a spin current source [48, 160, 164, 165], just like a HM layer. It has been found that the SOT efficiency of the single AFMs, such as PtMn, IrMn, is in the same order of Pt [160, 166, 167].

Interestingly, it was also found that there is an anisotropy of the ST-FMR results from epitaxial PtMn [48] and IrMn$_3$ [165] AFM films. For PtMn epitaxial films [48], the modulation of ST-FMR linewidth (MOD) is more significant along the *a*-axis than *c*-axis PtMn (10 nm)/Cu (1 nm)/Py (5 nm) samples as shown in the figures 24(a) and (b), indicating the large SOT efficiency in *a*-axis PtMn films. A further only-$V_S$ ST-FMR analysis method yields a SOT efficiency of ~0.048 for *c*-axis PtMn and ~0.089 for *a*-axis PtMn. For the case of epitaxial IrMn$_3$/Py bilayer devices [165], a similar anisotropic behaviour was found from ST-FMR measurements in the SOT efficiency with respect to the IrMn$_3$ crystal orientations. As shown in figure 24(c), the SOT efficiency is much larger (~0.2) in (001)-oriented single-crystalline IrMn$_3$ than that in either (111)-oriented or polycrystalline-oriented films. Moreover, after perpendicular field annealing, an enhanced SOT efficiency up to ~0.35 in (001)-oriented IrMn$_3$ thin films was demonstrated. Both works discussed that the observed anisotropic SOT efficiency in PtMn or IrMn$_3$ epitaxial films originates from the intrinsic SHE, which was further corroborated with the first-principles calculations.

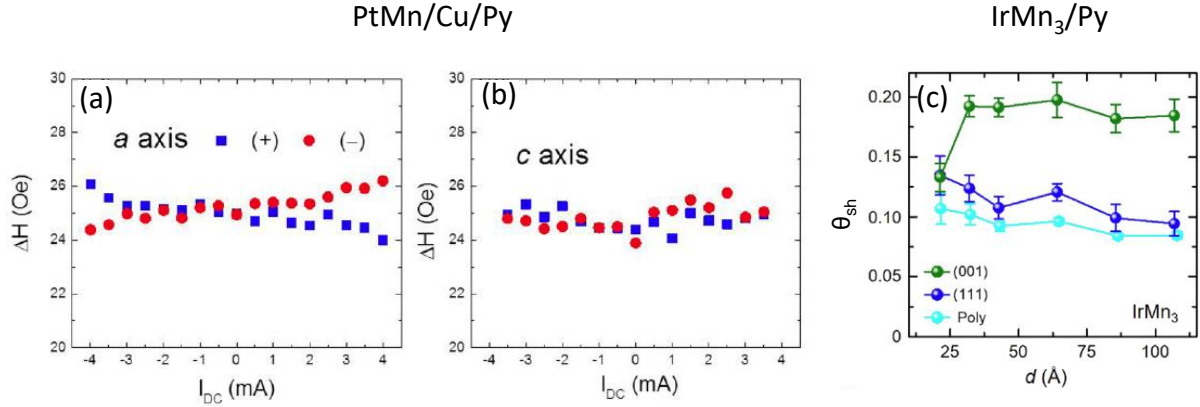

**Figure 24.** Measured resonance linewidth versus dc-bias current by ST-FMR measurements at 5 GHz for (a) *a*- and (b) *c*-axis PtMn (10 nm)/Cu (1 nm)/Py (5 nm). Reprinted with permission from [48]. Copyright (2015) by the American Physical Society. (c) SOT efficiency as a function of IrMn$_3$ thickness for different crystal orientation films, i.e. (001)-oriented (olive circles), (111)-oriented (blue circles), and polycrystalline-oriented (cyan circles). Reprinted from [165]. 2016 © The Authors, some rights reserved; exclusive licensee American Association for the Advancement of Science. Distributed under a Creative Commons Attribution NonCommercial License 4.0 (CC BY-NC).

**7.4. 2DEG in oxide materials**

Recent advances in film growth and synthesis techniques have enabled the growth of high quality oxide heterostructures with very smooth interfaces. This allows for engineering novel electronic properties at the oxide interfaces [168]. One of the central oxide material systems is the interface between two wide-band-gap insulators, SrTiO$_3$ (STO) and LaAlO$_3$ (LAO) as a 2DEG is formed at the interface and possesses exotic properties including superconductivity [169, 170] and magnetism [171-173]. Further, due to the broken interfacial inversion symmetry, the 2DEG confined in the vicinity of polar (LAO)/non-polar (STO) interface experiences a strong electric field directed perpendicular to the conduction plane [174]. Consequently, the LAO/STO interface possesses a strong Rashba SOC that leads to a strong coupling between the orbital and spin degrees of freedom. Due to the dielectric properties of STO, the strength of the Rashba SOC is tunable by applying an external gate voltage [175].

The presence of strong Rashba SOC in the *d*-bands of the 2DEG at the STO/LAO interface has been reported in various earlier studies [175-177]. Further, the Rashba SOC allows for generation of spin accumulation in the LAO/STO interface when a charge current flows in the 2DEG [138, 178]. Subsequently, the existence of current induced SOTs at the STO/LAO interface has been verified experimentally by angular dependent magnetoresistance measurements [179]. Recently, Wang *et al.*[180] have experimentally shown a giant room temperature charge-to-spin conversion efficiency (i.e. $\theta_{sh}$) in the STO/LAO/CoFeB structure by the ST-FMR technique shown in figure 25(a). The $\theta_{sh}$ value was estimated to be ~6.3 at room temperature denoted by squares in figure 25(b), which is almost two orders of magnitude larger than that in HMs, such as Pt and Ta [21, 25]. However, the $\theta_{sh}$ decreases rapidly as the temperature decreases to ~150 K. Finally, it becomes negligible at ~ 50 K. From the temperature dependent $\theta_{sh}$, it was suggested that inelastic tunneling via localized states, such as oxygen vacancies in the LAO band gap, accounts for the spin transmission through the LAO layer, as schematically shown in the inset of figure 25(b).

In addition to the ST-FMR measurements, a spin-to-charge conversion efficiency at the STO/LAO interface has been also reported by the spin pumping technique [181-183]. Additionally, a long spin diffusion length over 300 nm in the STO/LAO 2DEG channel has been reported [184, 185]. Therefore, the oxide materials, such as STO/LAO heterostructures, can enable potential applications in the oxide based spintronic devices.

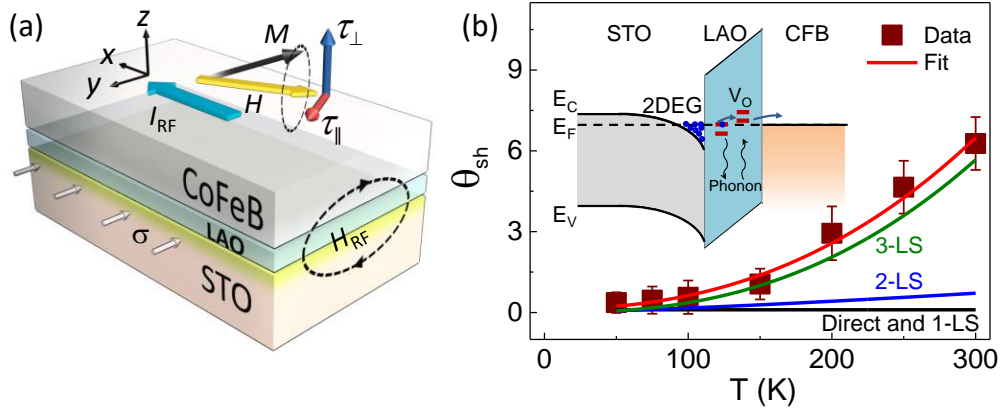

**Figure 25.** (a) Illustration of a STO/LAO/CoFeB ST-FMR device with the SOT induced magnetization dynamics. (b) SOT efficiency, $\theta_\parallel$, as a function of temperature (squares) and the fit (red line). The inset is the schematic of the spin-polarized electron inelastic tunneling process via the localized states (LS), such as oxygen vacancies, $V_O$. The fitting lines with the black, blue and green colours denote the contributions of the respective direct elastic tunneling & 1- localized state (LS), 2-LS and 3-LS chains. Reprinted with permission from [180]. Copyright (2017) American Chemical Society.

### 7.5. Nonmagnetic oxidized materials

In conventional HMs, the reported largest SOT efficiency is ~0.3 observed in $\beta$-phase W [35]. However, the $\alpha$-phase W exhibits a much smaller SOT efficiency. Therefore, the SOTs can be controlled via the HM microstructure engineering. Recently, it has been reported that the SOT efficiency can be significantly enhanced in a W/Py bilayer by oxygen incorporation in the W layer [186]. It was found that as the oxygen concentration increases, the grain size decreases, but the resistivity increases, in which $\beta$-phase W is well stabilized. Consequently, an enhanced SOT efficiency up to ~0.5 was achieved with an oxygen concentration of 12.1% compared to ~0.14 for zero oxygen incorporation. For higher oxygen incorporations up to 44%, the bulk properties further changed, however the SOT efficiency showed a very weak oxygen concentration dependence, suggesting that the mechanism responsible for the observed highly SOT efficiency may originate from the W (oxygen)/Py interface.

Recently, An *et al.* [187] found that by naturally oxidizing a light metal Cu with very weak SOC, the SOT efficiency in a Cu/Py bilayer was enhanced to be ~0.08, which is similar as Pt and more than two orders of magnitude larger than that in pure Cu. These results indicate that engineering materials with oxidization opens an intriguing path to enhance the SOT and realize efficient spintronic devices [188].

## 8. Summary and Perspectives

In this review, we first described the ST-FMR technique, including two ST-FMR setups and three ST-FMR analysis methods. As ST-FMR is a high frequency measurement in the GHz frequency range, several aspects should be carefully considered during the ST-FMR measurements and data analysis, such as the symmetric CPW design, impedance matching, rf power absorption and device current correction as well as possible contamination considerations (field-like torque from the interface effect and spin pumping). Then the latest SOT research progresses using ST-FMR are presented in diverse materials including normal HMs and alloys, TIs, 2D materials, interfaces with strong Rashba SOC, AFM materials, 2DEG in oxide materials as well as nonmagnetic oxidized materials. We find that the ST-FMR measurement is a powerful technique to determine the SOT efficiency including the recently observed out-of-plane damping-like SOT efficiency in a WTe$_2$ Weyl semimetal. We hope this review can further promote the ST-FMR technique in fundamental SOT researches as well as exploiting novel materials for future spin-based device applications.

In figure 26 we have summarized the representative SOT efficiencies in various materials as well as the power consumption for switching an unit magnetization of a FM layer which is proportional to $1/(\sigma \times (\theta_{sh})^2)$ [136]. Here we assume that the switching time is same in all the materials for simplicity. It shows that the TI materials generally show a larger SOT efficiency as well as low power due to the efficient charge-spin conversion in the spin-momentum-locked surface states compared to HMs. Some other materials, apart from HMs and TIs, also exhibit a high performance in terms of both SOT efficiency and power consumption and thus are worth for further studies.

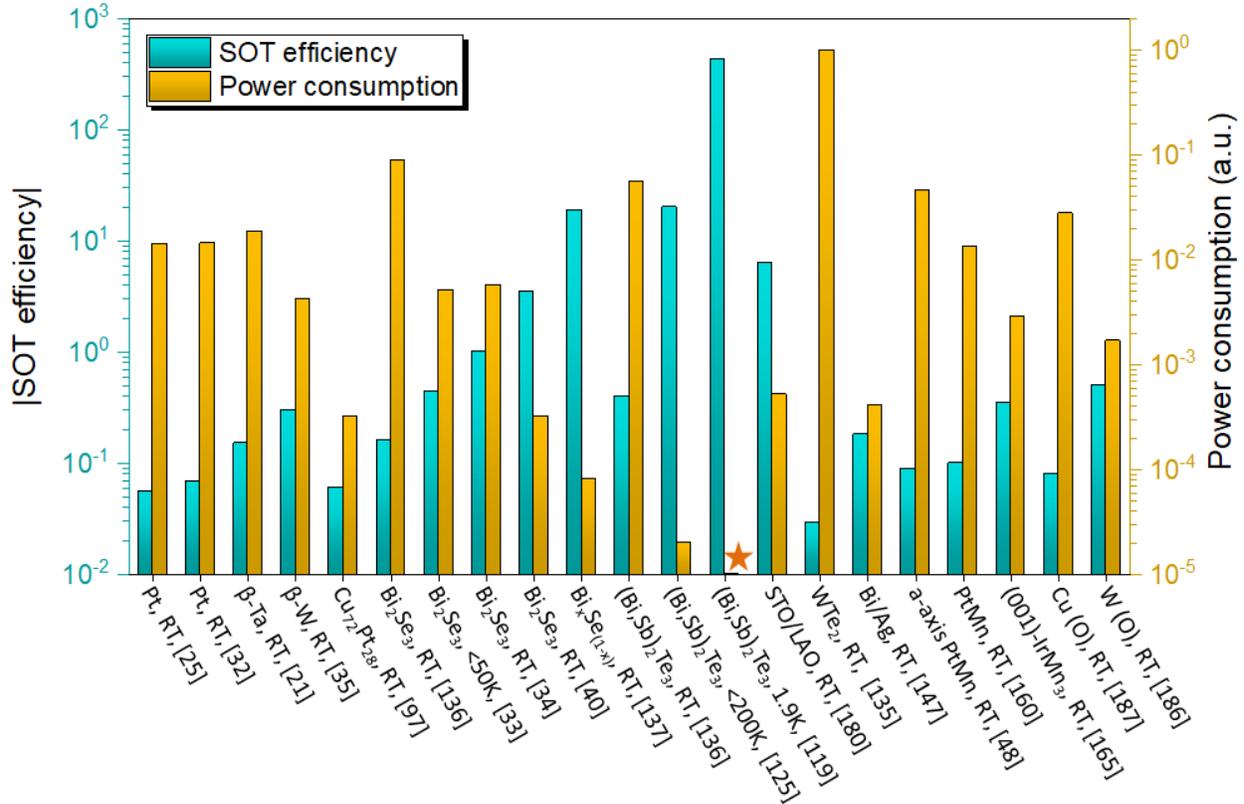

**Figure 26.** Summary of the SOT efficiencies (left *y*-axis) and normalized power consumption of a variety of materials (right *y*-axis). The *x*-axis denotes the materials with their corresponding measurement temperature and reference. RT means room temperature. The $Bi_xSe_{(1-x)}$ represents the film grown by the magnetron sputtering, the STO/LAO represents $SrTiO_3/LaAlO_3$ bilayer, and Cu (O) and W(O) represent the metal films with oxygen incorporation. The star represents an extremely small normalized power consumption value ($\approx 5.5 \times 10^{-8}$) for $(Bi,Sb)_2Te_3$ at a temperature of 1.9 K. The effective conductivity $\sigma$ (~0.17 $\mu\Omega^{-1}\cdot cm^{-1}$) of Bi 4 (nm)/Ag (8 nm) bilayer, used for the power consumption estimation, is calculated using a parallel circuit model with the given conductivity values of ~$2.1\times10^{-4}$ $\mu\Omega^{-1}\cdot cm^{-1}$ (for Bi) and ~0.25 $\mu\Omega^{-1}\cdot cm^{-1}$ (for Ag). The $\sigma$ (~0.0125 $\mu\Omega^{-1}\cdot cm^{-1}$) of Cu (O) layer, used for the power consumption estimation, is extracted from the figure 1(b) in Ref. [187] by using the given value of the pure Cu conductivity (~0.025 $\mu\Omega^{-1}\cdot cm^{-1}$).

Figure 27 shows the correlation of SOT efficiency and power consumption for different materials, obtained by reorganising and grouping the data in figure 26 based on their corresponding material system. Overall, we find that the TI materials (wrapped by the red dashed curve) shows a better performance in terms of the SOT efficiency and the power consumption even though the values are scattered due to the influence of conducting bulk. The HMs and their alloys (wrapped by the dashed blue curve) possess a similar orders of SOT efficiency and comparable power consumptions as the AFM systems but are inferior to the TI systems. Finally, the Weyl semimetal $WTe_2$ seems to show a higher power consumption based on the single report in Ref [135]. Since, Weyl semimetals are a very interesting topological materials, more research works are expected in the near future and one can obtain a more conclusive view about the performance of Weyl semimetals in the power consumption and SOT driven switching efficiency.

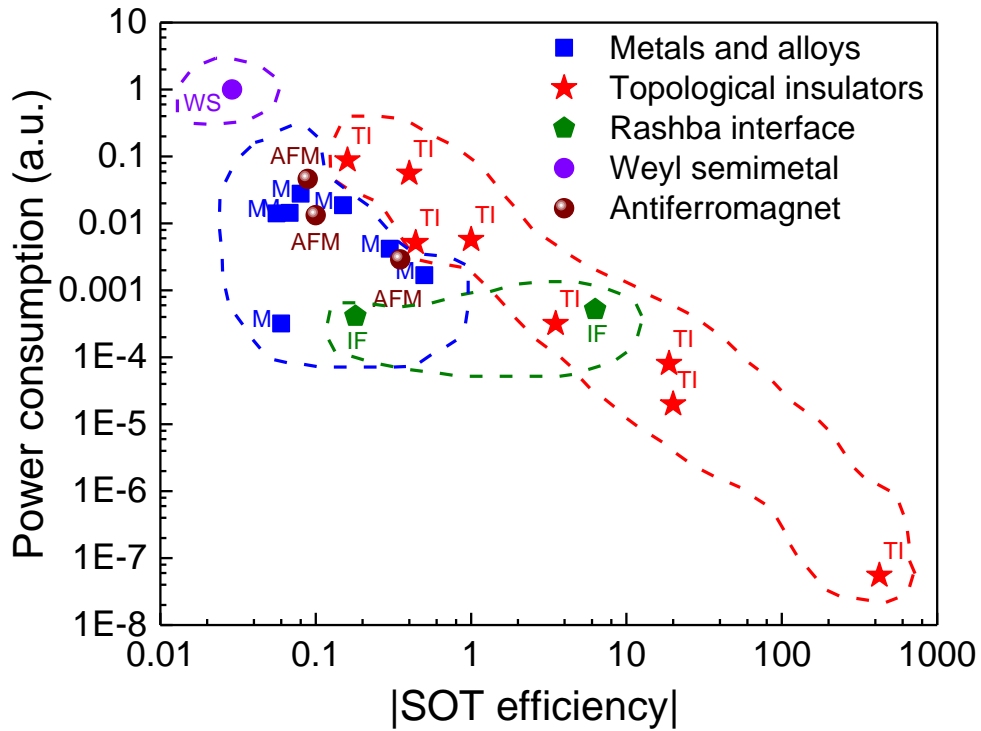

**Figure 27.** Summary of the normalized power consumption as a function of SOT efficiencies for a variety of materials. 'Metals and alloys' represent Pt, Ta, W, and CuPt alloy and nonmagnetic oxidized materials, 'Topological insulators' represent $Bi_2Se_3$ and $(Bi,Sb)_2Te_3$, 'Rashba interface' includes STO/LAO interface and Bi/Ag interface, 'Weyl semimetal' means $WTe_2$, and 'Antiferromagnet' includes PtMn and $IrMn_3$.

Moving towards to the real SOT applications, many reports recently have studied the SOT devices in the nanosecond (ns) or even picosecond (ps) regime [104, 108, 189-192]. A very high efficient SOT driven magnetization switching has been realized recently [191], where the switching energy is only about 60 fJ for each bit writing with the current ~120 µA and time of 1.2 ns. The energy consumption for each bit writing is much smaller than the values known in STT driven magnetization switching, which is between ~150 fJ to ~4 pJ [193-195]. Therefore, the SOT devices are promising and potential candidates for the future high efficient spintronic devices. Thus, the exploration of novel materials with even higher SOT efficiency will be of great importance for the future spin-based device applications and, for this purpose, the ST-FMR can be an easy and a powerful technique.

So far, we have seen that the origins of the SOT can be from the bulk, surface and/or interface in a variety of materials. Based upon the origins of the SOT, the material systems will show different thickness dependent behaviours. Figure 28 shows the schematic of the SOT efficiency as a function of thickness in representative TI material $Bi_2Se_3$, HM material Pt and Weyl semimetal $WTe_2$. For HMs (blue curve), the dominant mechanism for the SOT is SHE, which is a bulk property. Therefore, the SOT efficiency in HMs usually increases and then saturates as HMs thickness increases. The thickness corresponding to the saturation point of SOT efficiency is correlated to the spin diffusion length in HMs. For the TIs (red curve), the TSS serve as the main mechanism for the SOT and should have almost a constant SOT efficiency against the TI thickness as reported in Ref. [34] (also see figure 21(e)). However, as mentioned in Section 6.3, the contamination from the conducting bulk might be unavoidable. Consequently, the SOT efficiency shows much larger values as the TI thickness becomes thinner as TSS becomes dominant. As reported in WTe$_2$ [153], the out-of-plane damping-like SOT efficiency, due to a broken lateral mirror symmetry, remains almost constant with respect to the WTe$_2$ thickness, which is claimed to arise from interface mechanisms. The weak thickness dependence is illustrated by the green curve in figure 28. Therefore, the thickness dependent SOT efficiency (or SOT) measurements by ST-FMR or even other reliable techniques are very useful to get insight into the underlying physics of an emerging materials in the future.

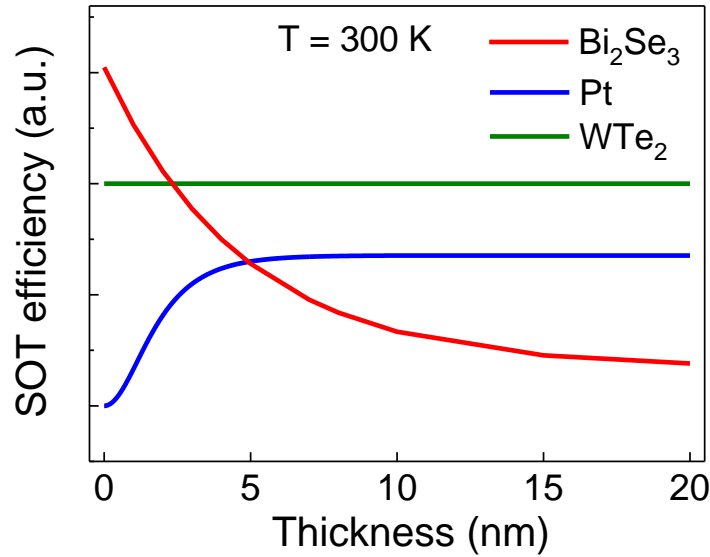

**Figure 28.** Schematic of the SOT efficiency as a function of thickness in representative materials Bi$_2$Se$_3$, Pt and WTe$_2$, replotted by using the data from figure 21(d) and (e), 8(a) and 23(e) in this review article, respectively.

The central physical effect used in the ST-FMR measurement can be further explored besides the GMR, tunneling magnetoresistance (TMR) or AMR rectification [196] in the present ST-FMR devices. The other magnetoresistance effects, such as the SMR, or Hall effects such as the anomalous Hall effect (AHE) and planar Hall effect (PHE), with both in-plane and out-of-plane measurement schemes, might also be developed in the near future to extend the ST-FMR technique to different material systems, especially for the materials with perpendicular magnetic anisotropy. Meanwhile, the ST-FMR analysis theory should be also modified accordingly. Subsequently, the SOT efficiency from ST-FMR measurements can provide a much better comparison with that from other measurements techniques. Good examples for this are the recent reports where the ST-FMR rectification signals were obtained based on the SMR effect in the Pt/yttrium iron garnet (YIG) magnetic insulator bilayer [197-199].

In addition, ST-FMR measurements on nontrivial topological materials, such as Weyl semimetals (WTe$_2$ [154] or MoTe$_2$ [200]), can be an future interesting topic. It would be significant discovery if one can observe the Fermi-arc surface states and/or Weyl node related topological phenomena using the ST-FMR technique. Apart from the above materials, the AFM materials are interesting to be further explored using ST-FMR measurements, as the AFMs are predicted to have a long spin coherence length due to the staggered spin order on an atomic scale and thus possess possible bulk-torque-like characteristics [201-203].

Besides the applications in the SOT studies, it has been demonstrated recently that TMR based ST-FMR devices have a very high efficiency in converting an rf frequency signal to dc signal, even better than the conventional Schottky diode [15, 204-206]. This high rectification efficiency of TMR based ST-FMR devices can be utilized for wireless energy harvesting that is becoming more relevant to the technologies such as internet of things (IoT).

**Acknowledgments**

This work was partially supported by the National Research Foundation (NRF), Prime Minister's Office, Singapore, under its Competitive Research Programme (CRP Award No. NRFCRP12-2013-01), NRF Industry-IHL Partnership (IIP) grant (R-263-000-C26-281), and A*STAR's Pharos Programme on Topological Insulators.